\documentclass{article}
\usepackage{arxiv}

\usepackage{graphicx} 
\usepackage{tabularx} 
\usepackage[table]{xcolor}
\usepackage[compact]{titlesec}
\usepackage[hidelinks]{hyperref}
\usepackage{amsmath}
\usepackage{amssymb}
\usepackage{mathtools}
\usepackage{amsthm}
\usepackage{comment}
\usepackage{float}
\usepackage{stfloats}

\usepackage{xcolor}

\usepackage{booktabs} 
\usepackage{siunitx}
\usepackage{multirow}
\usepackage{makecell}
\usepackage{subcaption}
\usepackage[capitalize,noabbrev]{cleveref}
\usepackage[linesnumbered, ruled]{algorithm2e}
\usepackage{array}
\newcolumntype{C}[1]{>{\centering\arraybackslash}p{#1}}
\usepackage{threeparttable}
\usepackage{pifont}  
\usepackage{tcolorbox}

\newtcolorbox{newpromptbox}[1][]{
    colback=gray!10,
    colframe=gray!50,
    boxsep=5pt,
    arc=4pt,
    fontupper=\ttfamily,
    #1
}

\usepackage{listings}
\usepackage{xcolor}
\definecolor{lightgray}{rgb}{0.95,0.95,0.95}

\newcommand{\plaintext}{\hat{m}}
\newcommand{\share}{\bar{m}}

\newcommand{\subject}[1]{\noindent\textbf{#1}}
\newcommand{\subsubject}[1]{\noindent\textit{\underline{#1}}}

\newcommand\malurl[1]{\href{notalink}{{\nolinkurl{#1}}}}
\newcounter{finding}
\newcommand{\ignore}[1]{}

\usepackage{todonotes}

\newcolumntype{L}[1]{>{\raggedright\let\newline\\\arraybackslash\hspace{0pt}}m{#1}}
\newcolumntype{C}[1]{>{\centering\let\newline\\\arraybackslash\hspace{0pt}}m{#1}}
\newcolumntype{R}[1]{>{\raggedleft\let\newline\\\arraybackslash\hspace{0pt}}m{#1}}

\newcommand{\protocol}{\textit{\textsf{Pepper}}}

\newcommand{\pepper}{\texttt{{Pepper}}}

\usepackage{amsthm}

\theoremstyle{definition}
\newtheorem{definition}{Definition} 

\usepackage{enumitem}

\usepackage{cite}

\title{Suan: Scalable and Efficient Metadata-Private Anonymous Broadcast}
\title{Pepper: Scalable and Efficient Metadata-Private Anonymous Broadcast}
\title{Pepper: Scalable and High-bandwidth Anonymous Broadcast with Cryptographic Privacy}
\title{Pepper: High-bandwidth and Deployable Anonymous Broadcast with Cryptographic Privacy}
\title{Pepper: High-bandwidth and Scalable Anonymous Broadcast with Cryptographic Privacy}
\title{Pepper: High-bandwidth and Scalable Anonymous Broadcast with Cryptographic Privacy\thanks{An abridged version of this work has been accepted for publication in IFIP SEC 2026.}}

\newcommand{\ustc}{USTC}
\author{
  Chenghao Li \\
  \ustc 
  \And
  Haoyuan Wang \\
  \ustc
  \And
  Xianghang Mi\\
  Monash University
}
\date{}

\begin{document}

\thispagestyle{plain}
\pagestyle{plain}

\maketitle

\begin{abstract}

We present \texttt{Pepper}, a high-bandwidth anonymous broadcast protocol that provides cryptographic sender anonymity against global adversaries. \texttt{Pepper} builds on a two-server DC-net architecture but introduces three key innovations: a self-contained anonymous registration subprotocol using verifiable distributed point functions, support for batch messaging via distributed multi-point functions, and a lightweight access control mechanism based on secret-shared proofs. Unlike prior systems, \texttt{Pepper} eliminates the need for external dialing services and allows each broadcaster to send multiple messages per epoch with a single audit, significantly improving throughput for large data transfers. Our implementation demonstrates that \texttt{Pepper} achieves millisecond-level registration audits, scales efficiently to thousands of channels, and delivers 1.2--20$\times$ higher effective messaging rates than state-of-the-art alternatives. Furthermore, \texttt{Pepper} is designed for practical deployment, with natural compatibility for co-deployment alongside Tor and federated social networks.
   
\end{abstract}
\section{Introduction}
\label{sec:intro}
Anonymous broadcast enables participants to send messages to the public while concealing the identity of the sender from all observers, including a global network adversary. This capability is essential for privacy-sensitive applications such as whistleblowing, censorship-resistant publishing, and secure communication under surveillance. Traditional anonymity systems like Tor~\cite{dingledine2004tor} provide low-latency routing but remain vulnerable to traffic-correlation attacks by global adversaries~\cite{overlier2006locating, hopper2010much, kwon2015circuit}. To achieve stronger, cryptographically guaranteed sender anonymity, a line of research has developed \emph{anonymous broadcast protocols} (ABPs) based on architectures such as Dining Cryptographers networks (DC-nets)~\cite{chaum1988dining} and mix-nets~\cite{chaum1981untraceable}. These protocols guarantee that even an adversary who passively observes all network traffic and controls a subset of servers cannot link a published message to its originator beyond a random guess within an anonymity set.

Despite significant advances, practical deployment of high-bandwidth anonymous broadcast faces several persistent challenges. Modern DC-net-based systems like Spectrum~\cite{newman2022spectrum} and Express~\cite{eskandarian2021express} employ distributed point functions (DPFs)~\cite{gilboa2014distributed,boyle2016function} to compress client-to-server communication and enable efficient server-side auditing. However, these systems exhibit critical limitations: they rely on an unspecified external ``dialing'' or registration system for channel allocation, lack native support for batch messaging (limiting throughput for large data transfers), and often impose complex auditing mechanisms that hamper secure implementation and scalability. Other approaches, such as Blinder~\cite{abraham2020blinder}, employ information-theoretic secret sharing to tolerate server faults but require an honest majority of servers—a less practical trust assumption—and incur high communication overhead. Furthermore, prior works have not fully explored compatibility with existing large-scale decentralized infrastructures, such as Tor or federated social networks (e.g., Mastodon~\cite{mastodon_project}), which could provide natural sources of cover traffic and server resources.

In this paper, we present {\pepper}, a high-bandwidth anonymous broadcast system that addresses these limitations through a novel, integrated design. {\pepper} builds upon a two-server DC-net architecture but introduces three key innovations:

\begin{itemize}[leftmargin=*, label=\textbullet]
    \item \textbf{Self-contained anonymous registration.} {\pepper} specifies a full registration subprotocol that allows a broadcaster to anonymously acquire one or more message channels without relying on any external dialing service. The protocol uses \emph{verifiable distributed point functions} (VDPFs)~\cite{de2022lightweight} to let aggregation servers jointly verify registration requests without learning the target channels under registration, ensuring both anonymity and robustness against malformed inputs.

    \item \textbf{Batch messaging with efficient auditing.} Unlike prior DC-net systems that restrict each broadcaster to one message per epoch, {\pepper} supports \emph{batch messaging}, allowing a single broadcaster request to carry up to $t$ messages to distinct channels. This is enabled by recent constructions of \emph{distributed multi-point functions} (DMPFs)~\cite{boyle2025improved} and their verifiable variants (VDMPFs). Crucially, \pepper's auditing mechanism verifies an entire batch in a single shot, requiring only one proof exchange regardless of $t$, thereby amortizing verification overhead and achieving high effective throughput.

    \item \textbf{Lightweight access control via secret-shared proofs.} For messaging, {\pepper} integrates a Schnorr Proof over Secret Shares (SPoSS) primitive~\cite{servan2023private} to enforce that only the owner of a registered channel can write to it, while allowing cover-traffic clients to generate dummy writes using publicly known ``phantom channels.'' This approach provides robust access control with minimal server-side computation, avoiding the expensive per-channel exponentiations required by earlier designs like Spectrum.
\end{itemize}

These technical contributions make {\pepper}uniquely suited for high-bandwidth applications such as anonymous file sharing or media dissemination. Our implementation and evaluation demonstrate that {\pepper} achieves:
\begin{itemize}[leftmargin=*, label=\textbullet]
    \item \textit{Low auditing latency:} Registration audits complete in milliseconds even for 16K channels, and batch-messaging audits scale efficiently with the batch size $t$.
    \item \textit{High effective throughput:} For batch size $t=8$, {\pepper} delivers $1.2\times$ to $5.0\times$ higher messaging rates than PACLs~\cite{servan2023private} and $13.6\times$ to $20\times$ higher rates than Spectrum across a range of channel counts.
    \item \textit{Practical compatibility:} {\pepper}'s architecture naturally aligns with existing infrastructures; aggregation servers can be co-deployed as Tor relays, and its directory service can integrate with Tor's directory authorities. Moreover, federated social network users can seamlessly provide cover traffic.
\end{itemize}

In summary, this paper makes the following contributions:
\begin{itemize}[leftmargin=*, label=\textbullet]
    \item We design and specify {\pepper}, an anonymous broadcast protocol that integrates a full anonymous registration subprotocol with support for efficient batch messaging, eliminating the need for external dialing systems.
    \item We construct a lightweight yet robust access control mechanism that combines VDMPFs with secret-shared proofs (SPoSS), enabling efficient verification of batch writes and seamless support for cover traffic.
    \item We implement {\pepper} in approximately 8,000 lines of Go and 2,900 lines of C/C++, and evaluate it against state-of-the-art baselines (Spectrum, Express, and PACLs), demonstrating superior throughput and scalability for high-bandwidth broadcasts. The code is open source and available at \url{https://github.com/ChaseSecurity/Pepper}.
    \item We show how {\pepper} can be practically deployed alongside existing decentralized networks like Tor and Mastodon, leveraging their infrastructure for cover traffic and server resources.
\end{itemize}

\subject{Paper structure.} Section~\ref{sec:background} reviews background on anonymous broadcast and related work. Section~\ref{sec:overview} presents {\pepper}'s threat model, design goals, and high-level architecture. Section~\ref{sec:system} details the registration and messaging subprotocols. Section ~\ref{sec:analysis} analyzes the theoretical efficiency and security of the system. Section~\ref{sec:eval} provides implementation details and comprehensive evaluation. Section~\ref{sec:discuss} discusses limitations and future work, and Section~\ref{sec:conclusion} concludes.
\section{Background and Related Works}
\label{sec:background}

In this section, we introduce the foundational concepts and settings necessary to contextualize and compare anonymous broadcast protocols.

\subject{Motivation and Definition of Anonymous Broadcast.} Traditional network communication systems, such as TLS-based client-server communication, do not inherently guarantee anonymity. Specifically, the sender's metadata, including IP addresses, is visible not only to the recipient but also to any passive traffic observer, whether local or global. This lack of anonymity is problematic in scenarios where privacy is paramount, such as whistleblowing or communication under oppressive censorship. To address this gap, systems like the Tor network (onion routing)~\cite{dingledine2004tor} have been developed and widely deployed. However, despite providing anonymity against local adversaries, Tor remains vulnerable to global adversaries capable of de-anonymizing senders through various traffic correlation attacks~\cite{overlier2006locating, hopper2010much, kwon2015circuit}.

To achieve sender anonymity against global adversaries, various protocols have been proposed. While some focus on anonymous peer-to-peer messaging, most are designed for anonymous broadcast, and broadcast protocols can be easily adapted to peer-to-peer anonymous messaging, e.g., receiver anonymity can be achieved via private information retrieval~\cite{kwon2017atom}. 

Formally, and in alignment with prior work, we define \textit{anonymous broadcast} as a cryptographic protocol that guarantees sender anonymity (metadata hiding) in the presence of a global adversary capable of passively observing all network traffic, controlling a subset of servers and any number of clients providing cover traffic. Some anonymous broadcast protocols (ABPs)~\cite{van2015vuvuzela, tyagi2017stadium, lazar2018karaoke} offer differential privacy guarantees, trading quantifiable privacy leakage for efficiency. Others, including {\pepper}, provide cryptographic privacy, ensuring zero metadata leakage under a well-defined threat model.

Among ABPs offering cryptographic privacy, protocols can be further categorized by whether their underlying cryptographic primitives are built upon information-theoretic cryptography or computational cryptography. Information-theoretic protocols~\cite{abraham2020blinder} are secure against adversaries with unlimited computational power but are often very costly and less practical. Computational cryptography-based protocols~\cite{newman2022spectrum, eskandarian2021express}, on the other hand, rely on the hardness of specific mathematical problems, providing conditional security. {\pepper} falls into the latter category, leveraging computational cryptography.

A typical ABP involves several key concepts. The \textit{anonymity set} refers to the total number of honest clients $N$, encompassing both broadcasters and cover-traffic participants that are honest. The fundamental property of an ABP is that neither a passive global adversary nor any participating client or server can correlate a broadcasted message with its respective broadcaster within the anonymity set. Naturally, larger anonymity sets provide stronger anonymity. In real-world deployments, cover traffic participants can be recruited from diverse sources, such as Tor clients or users of decentralized social networks like Mastodon. As with other distributed systems, ABPs must contend with Byzantine nodes, which may include malicious clients or servers controlled by the adversary.

To detect or even prevent malicious clients and servers, an auditing subprotocol is typically introduced. The complexity and overhead of this subprotocol depend on the auditing goals. Lightweight auditing mechanisms aim to detect and, in some cases, attribute a protocol distruption to specific partiews~\cite{newman2022spectrum}. Upon detecting a disruption, the given communication round is aborted, and the responsible party, if identified, is excluded. However, lightweight auditing cannot prevent denial-of-service attacks. To (partially) address such attacks, some protocols, such as Blinder~\cite{abraham2020blinder}, employ more robust auditing mechanisms capable of recovering from disruptions, provided the proportion of malicious nodes remains within defined limits (e.g., in Blinder, $t < n/4$, where $t$ is the number of malicious servers and $n$ is the total number of servers).

Next, we explore the key architectures of ABPs and provide a detailed comparison of {\pepper} with closely related protocols.


\subject{DC-net based Anonymous Broadcast.} The Dining Cryptographers network (DC-net), introduced by Chaum~\cite{chaum1988dining}, enables multiple parties to compute a function of their private inputs without revealing the inputs themselves. The original DC-net scheme is fully distributed. In a naive distributed DC-net setup, only one client is permitted to send a non-zero message $m_i = \plaintext$, while all other clients send zero messages ($m_i = 0$). Each epoch consists of two stages. First, every pair of clients privately negotiates a shared secret $r_{i,j}$. In the second stage, each client computes a local aggregation of their message $m_i$ with the $N-1$ negotiated secrets as ${\share}_i = m_i \oplus \bigoplus_j r_{i,j}$ and publishes the aggregated result $\share_i$. The public aggregation of all local results reveals the true message $\plaintext$ while preserving sender anonymity. However, this approach incurs a high communication cost that grows quadratically with the number of participating clients.

To address this inefficiency, later protocols introduced centralized DC-net architectures, where aggregation and result publishing are delegated to a small set of designated servers, namely \textit{aggregation servers}. This server-based DC-net design has become the foundation of many anonymous broadcast protocols (ABPs), including \pepper, due to its improved communication efficiency and scalability. In a basic two-server instantiation, each client $i$ splits their message $m_i$ into two additive shares. For example, the client samples a random pad $r_i$, sends $r_i$ to server A, and $r_i \oplus m_i$ to server B. Each server $j$ aggregates the received shares $\share_{i}^{j}$ from all clients to compute a local aggregation $\share^{j}_{agg} = \bigoplus_{i=1}^{N} \share_{i}^{j}$. By exchanging and combining these local aggregations, the servers reconstruct the final message $\plaintext$. Unlike the distributed DC-net, this centralized approach reduces the overall communication overhead to linear complexity with respect to the number of clients. 

However, similar to its distributed counterpart, only a single non-zero message is allowed per epoch, resulting in low throughput that is impractical for many real-world applications, e.g., anonymous microblogging. 
To enhance throughput, recent advancements~\cite{newman2022spectrum,abraham2020blinder} have introduced multi-channel server-based DC-nets, where servers maintain $L$ channels, allowing up to $L$ messages to be broadcasted per epoch, with each message attached to a distinct channel. 

Naive server-based DC-nets are susceptible to disruption attacks, where a single malicious client can inject malformed shares to impede successful message reconstruction (jamming). Modern DC-net variants address this vulnerability by incorporating novel access control mechanisms to audit writes before aggregation, rejecting malformed or unauthorized requests while preserving index privacy. Practical systems also batch writes into fixed-time epochs, use channels to separate broadcasters, and implement blame or accountability protocols to evict byzantine or malicious participators. These enhancements significantly improve robustness and throughput, making server-based DC-net systems increasingly viable for large-scale anonymous communication.

\subject{Mixnet Based Anonymous Broadcast.} In contrast to DC-net-based ABPs, another class of anonymous broadcast protocols is based on the mixnet architecture~\cite{chaum1981untraceable}. A basic mixnet consists of layers of servers, where each server shuffles the received messages randomly before forwarding them to the next layer. While this design is communication-efficient, it is vulnerable to various traffic analysis attacks, such as deliberate packet dropping or delaying. 

To address these vulnerabilities, modern mixnets incorporate secure and verifiable shuffling protocols, often leveraging computationally expensive multi-party computation (MPC) techniques to ensure that servers shuffle messages honestly and in a verifiably random order. However, these secure shuffling protocols introduce significant overhead in both communication and computation. Consequently, compared to DC-net-based ABPs, mixnet-based ABPs~\cite{chaum2016cmix, kwon2017atom, tyagi2017stadium, cho2024rabbit, langowski2022trellis} typically require a larger number of servers, stronger and less practical trust assumptions (e.g., assuming majority of honest servers), and additional communication rounds, particularly among shuffling servers.

\begin{table}
    \centering
    \caption{Qualitative comparison between \protocol with leading anonymous broadcast protocols based upon centralized DC-net (client-server DC-net).}
    \label{tab:comparing_sota_abps}
    \begin{threeparttable}
        \begin{tabular}{cccccc}
            \toprule
            Protocol & Per C-Comp~\tnote{1} & Per S-Comp~\tnote{1} & C-S Comm~\tnote{1} & S-S Comm~\tnote{1}  & Dialing\tnote{2}\\
            \midrule
            Blinder~\cite{abraham2020blinder} 
                & $\sqrt{N} \cdot |m|$ 
                & $N^2 \cdot |m|$ 
                & $\sqrt{N} \cdot |m|$ 
                & $N \cdot |m|$ 
                & N/A\\
            Spectrum~\cite{newman2022spectrum} I~\tnote{3} 

                   & $\log L + |m|$ 
                & $N \cdot L \cdot |m|$
                & $\log L + |m|$
                & $L \cdot |m| + N$
                & \ding{56}\\
            Spectrum~\cite{newman2022spectrum} II~\tnote{3} 
                & $\sqrt{L} + |m|$
                & $N \cdot L \cdot |m|$ 
                & $ \sqrt{L} + |m|$
                & $L \cdot |m| + N$
                & \ding{56}\\
            Express~\cite{eskandarian2021express} 
                & $\log L + |m|$ 
                & $N \cdot L \cdot |m|$
                & $\log L + |m|$
                & $L \cdot |m| + N$
                & \ding{56}\\
            \bottomrule
            \pepper 
                & $t \log L + t|m|$ 
                & $N \cdot L \cdot |m|$
                & $t^2 \log L + t |m|$
                & $L \cdot |m| + N$
                & \ding{52}\\
            \bottomrule
        \end{tabular}
        \begin{tablenotes}
            \item [1] $n$ denotes the number of servers, while the number of clients is represented as $N$; $t$ is the message batch size (number of messages per client request for \protocol). $O(\cdot)$ is omitted for clarity. Also, we assume the security parameter $\lambda$ and the number of servers are pre-configured constants. 
            \item [2] Whether a dialing (registration) subprotocol is required and specified as part of the ABP. \textit{N/A} denotes the respective ABP doesn't require a dialing subprotocol.
            \item [3] Spectrum~I refers to the 2-server Spectrum protocol, while Spectrum~II denotes the multi-server setting.
        \end{tablenotes}
    \end{threeparttable}
\end{table}

\subject{Comparison between {\pepper} and Representative Counterparts.} As {\pepper} adopts the server-based DC-net architecture, we focus our comparison on DC-net counterparts. Given the extensive works of DC-net ABPs~\cite{wolinsky2012dissent, corrigan2015riposte}, we focus our comparison on recent protocols that utilize advancements in cryptographic primitives and achieve state-of-the-art performance in terms of  communication and computation. These include Spectrum~\cite{newman2022spectrum}, Blinder~\cite{abraham2020blinder}, and Express~\cite{eskandarian2021express}. All these protocols provide cryptographic privacy, ensuring no leakage of sender metadata. 

Spectrum operates under a practical anytrust threat model, assuming at least one honest aggregation server. In each epoch, broadcasters generate additive shares of their messages, with the number of shares equal to the number of aggregation servers. Servers perform channel-wise aggregation locally and exchange aggregated shares. To reduce client-server communication, Spectrum employs distributed point functions (DPFs)~\cite{gilboa2014distributed, boyle2016function}. Despite its practical threat model and focus on high-bandwidth anonymous broadcasting, Spectrum has notable limitations. It relies on the availability of an anonymous dialing system for a broadcasters to register channels and her public keys, but fails to specify the dialing subprotocol, leaving this dependency unresolved. Additionally, Spectrum is inherently designed for one channel per broadcaster, rendering its throughput per broadcaster infeasible for high-bandwidth applications, e.g., broadcasting a 1GB file costs over 13 hours.

Express~\cite{eskandarian2021express}, originally designed for anonymous whistleblowing, can be adapted for anonymous broadcast. Like Spectrum, it employs a two-server DC-net and uses DPFs to compress client-server communication. For auditing, Express incorporates a large virtual address space to prevent write collisions and zero-knowledge proofs to verify input formats. However, it shares Spectrum's reliance on an external anonymous dialing system. Furthermore, Express is optimized for low-bandwidth scenarios, with evaluations limited to messages of up to 32 KB. Its large virtual address space mitigates targeted client disruption but cannot prevent collusion between malicious clients and servers, which could leak virtual addresses.

Blinder~\cite{abraham2020blinder} differs from Spectrum and Express by using Shamir's threshold secret sharing~\cite{shamir1979share}, enabling recovery from disruptions caused by malicious servers. However, this approach requires an honest majority assumption, where at most one-fourth of the servers can be malicious, which is less practical than the anytrust model. Additionally, Shamir's secret sharing necessitates complex and costly auditing to verify that client shares align with the same polynomial.

In summary, all three counterparts exhibit one or more of the following limitations: (1) reliance on an external anonymous dialing system; (2) complex and resource-intensive auditing protocols; (3) insufficient throughput for high-bandwidth applications; and (4) unclear compatibility with existing systems such as Tor or decentralized social networks, which none of these studies have extensively explored or demonstrated.

\subsubject{Efficiency Analysis.} Table \ref{tab:comparing_sota_abps} provides a comparative analysis of the proposed system (\pepper) against three state-of-the-art anonymous broadcast protocols: Spectrum, Blinder, and Express. The comparison focuses on key metrics, including per-client computation complexity per epoch (\textit{C-Comp}), per-server computation complexity per epoch (\textit{S-Comp}), client-to-server communication size per client per epoch (\textit{C-S Comm}), server-to-server communication size per aggregation server per epoch (\textit{S-S Comm}), and whether the protocol specifies a dialing subprotocol (\textit{Dialing}).

Among the protocols, {\pepper} is the only system that explicitly specifies a dialing (registration) subprotocol. This feature makes {\pepper} self-contained, eliminating the need for an external anonymous registration system, which is a requirement for Spectrum and Express. While Blinder does not require a dialing system, it relies on maintaining a large channel space proportional to the number of clients, which results in reduced efficiency in terms of communication size and computation complexity.


Furthermore, {\pepper} supports batch messaging, allowing up to $t$ messages per broadcaster per epoch. This capability improves end-to-end throughput over Spectrum and Express, as demonstrated in the throughput evaluation (Section~\ref{subsec:throughput}), but it also introduces a higher per-epoch cost in client-side computation and client-to-server communication. To keep such batch requests compact and secret-shared, {\pepper} relies on distributed multi-point functions (DMPFs), which generalize DPFs from a single non-zero point to multiple non-zero points~\cite{boyle2025improved}. A DMPF represents a sparse vector with up to $t$ non-zero points as a single pair of function shares, rather than as $t$ independent DPF instances, and therefore matches the DPF-style request format used by DC-net based anonymous broadcast. This allows {\pepper} to batch multiple channel writes while retaining comparable server-side computation and server-to-server communication. The Big-State construction of Boyle et al.~\cite{boyle2025improved} is one representative DMPF construction: its key size grows with the public sparsity bound $t$, so adapting DMPFs for {\pepper}-style batch writes does not make the cost independent of the number of supported points. Consequently, protocols that wish to hide the actual number of real messages must still pad each request to a public maximum batch size and smaller encodings can improve communication but directly reveal how many points were encoded.

Besides, {\pepper} also supports servers to validate submitted function shares without learning the target channels. This corresponding verification layer comes from verifiable DPFs. Boyle et al.'s FSS extensions provide protocols for checking whether user-supplied FSS keys are consistent with some function in the claimed family~\cite{boyle2016function}. De Castro and Polychroniadou subsequently show that lightweight, maliciously secure VDPFs can verify, using only symmetric-key operations, whether two DPF key shares encode a valid point function without revealing the target point~\cite{de2022lightweight}. For single-point writes, this ensures that the submitted key shares jointly define a DPF whose evaluations combine to a vector with a single non-zero point. For multi-point writes, the same verifiable-FSS interface can be applied to DMPFs: key generation produces DMPF shares together with verification material, evaluation returns both an output share and a verification token, and servers check these tokens before accepting the evaluated shares~\cite{boyle2016function,de2022lightweight}. This yields verifiable DMPFs whose checks certify that the submitted shares encode a valid sparse function under the public bound $t$, without revealing the non-zero positions~\cite{boyle2016function,de2022lightweight,boyle2025improved}.

\section{System Overview}
\label{sec:overview}

Upon related concepts discussed in Section~\ref{sec:background}, This section provides a high-level overview of {\pepper}, with regards to its threat model and design goals (Section~\ref{subsec:threat_model}), architecture (Section~\ref{subsec:architecture}), and main components (Section~\ref{subsec:regi_overview} and Section~\ref{subsec:msg_overview}). 

\subsection{Threat Model and Security Goals}
\label{subsec:threat_model}
Next, we elaborate on the threat model of \pepper, focusing on the adversary's capabilities and the trust assumptions on participating entities.

Pepper assumes a global adversary capable of passively observing communication traffic across the entire network and performing advanced traffic analysis attacks. However, communication between broadcasters and servers is secured using TLS-like protocols, ensuring that the adversary cannot access the plaintext of encrypted communication unless it controls the aggregation server. This aligns with the security guarantees of modern cryptographic protocols.

Regarding aggregation servers, Pepper adopts a flexible any-trust assumption, requiring at least one honest server among the aggregation server set to maintain its anonymity guarantees. The remaining servers may be controlled by the adversary and may collude to de-anonymize broadcasters. This assumption is consistent with the design of many anonymous communication systems.

Pepper also relies on distributed clients to provide cover traffic. Some of these clients may be malicious, controlled by the adversary, or behave in a Byzantine manner, such as dropping out during an epoch, selectively sending cover traffic, or generating malformed messages. Despite this, the system assumes that a subset of cover-traffic clients remains honest and uncompromised, forming an anonymity set for broadcasters.

\subject{Design Goals.} The design of \pepper is guided by the following key objectives:

\subsubject{Anonymity.} Building on the threat model outlined earlier, {\pepper} is designed to ensure the anonymity of broadcasters, which is its foremost design goal. Specifically, given a published message, the likelihood of attributing the message to its true broadcaster should be indistinguishable from a random guess within the anonymity set, which includes both the broadcaster and a substantial number of honest cover-traffic clients.

\subsubject{High Throughput.} Beyond anonymity, {\pepper} aims to achieve a level of throughput that surpasses prior works as well as being practical for real-world applications, such as broadcasting large files or short video content on social media platforms. This focus on performance ensures that {\pepper} can support modern communication demands.

\subsubject{Scalability and Fault Tolerance.} \pepper is architected for horizontal scalability, enabling the system to handle increased throughput by deploying additional servers. The default configuration requires only a pair of aggregation servers, simplifying deployment and reducing operational overhead. In practical deployments, such as integration with the Tor network, multiple aggregation cohorts (each consisting of two servers) can operate concurrently. Broadcasters can dynamically select an appropriate cohort by consulting a public bulletin, ensuring flexibility and scalability.

Fault tolerance is another critical aspect of \pepper's design. While both aggregation servers and cover-traffic clients may exhibit Byzantine behavior, {\pepper} incorporates mechanisms to detect protocol deviations. Recovery is supported for client deviations, allowing the system to maintain high throughput. However, recovery from server deviations is not prioritized to avoid compromising performance. Instead, upon detecting a server deviation, the broadcaster can abandon the current epoch and select a new cohort to initiate a fresh broadcast. The presence of many aggregation cohorts in real-world deployments ensures robust fault tolerance.

\subsubject{Compatibility with Existing Decentralized Infrastructures.} Unlike prior works, {\pepper} emphasizes compatibility with existing decentralized infrastructures, such as the Tor network and Mastodon, a prominent decentralized social network. This compatibility facilitates rapid adoption and deployment by leveraging existing infrastructure to scale both the cover-traffic client base and the aggregation server cohorts. With tens of thousands of servers and millions of users, these infrastructures provide a solid foundation for achieving \pepper's goals of high throughput and fault tolerance. Detailed discussions on \pepper's compatibility with these systems are provided in Section~\ref{subsec:extension}.



\subsubject{Non-Goals.} {\pepper} does not aim to address the following: (1) Denial-of-service attacks originating from a huge volume of malicious clients or malicious ISPs and Internet exchange points; (2) Intersection attacks that exploit participation patterns wherein the anonymity set is small; (3) Vulnerabilities at the network or transport layer, such as weaknesses in TLS.

\subject{Cryptographic Assumptions.} The security of {\pepper} relies on well-established cryptographic primitives, including the hardness of the discrete logarithm problem, the decision Diffie-Hellman problem, and the existence of secure hash functions and pseudorandom generators.

\subsection{System Architecture}
\label{subsec:architecture}

To achieve the design goals outlined earlier, {\pepper} incorporates four distinct roles and operates through a two-phase protocol, as detailed below.

\subject{Participating Roles.} The Pepper system involves the following four roles: (1) \textit{A cohort of aggregation servers}; (2) \textit{A directory service}; (3) \textit{Cover-traffic clients}; and (4) \textit{Broadcasters}.

The cohort of aggregation servers is central to the system, responsible for aggregating and publishing messages. By default, each cohort consists of two servers, a configuration that enables the use of highly efficient cryptographic primitives, such as two-server distributed point function schemes. In real-world deployments, volunteer aggregation servers from diver sources(e.g., Tor relays or Mastodon servers) can form multiple cohorts. However, only a single cohort is required to anonymously broadcast a message, ensuring horizontal scalability since adding more servers increases the system capacity. Under the any-trust assumption, where at least one server in a cohort must be honest, the process of forming cohorts should either be managed by a trusted entity (e.g., a trusted directory service) or follow a verifiable random process. The specifics of cohort construction are beyond the scope of {\pepper}.

Once cohorts are established, the directory service maintains an up-to-date registry of active cohorts and provides a lookup service. Both broadcasters and cover-traffic clients periodically query the directory service to identify active cohorts, a process that does not compromise broadcaster anonymity. Additionally, the directory service collects and disseminates cohort statistics, such as the average number of clients per epoch, as reported by aggregation servers. These statistics serve as valuable indicators for broadcasters when selecting cohorts, with higher client participation per epoch typically indicating stronger anonymity guarantees.

The remaining roles—broadcasters and cover-traffic clients—are straightforward. Broadcasters are responsible for publishing messages, while cover-traffic clients generate dummy traffic to enhance anonymity. Notably, each cohort can support broadcasters up to the number of available channels per epoch. Furthermore, {\pepper} supports \textit{batch messaging}, which allows broadcasters to register multiple channels and publish messages to all these channels in a single epoch, thereby achieving higher throughput per broadcaster.

With the roles defined, we now summarize their interactions, which collectively enable the anonymous delivery of messages from broadcasters.

\subject{Protocol Phases.} From the broadcaster's perspective, Pepper operates through two sequential subprotocols. Once a cohort of aggregation servers is selected, the broadcaster (along with cover-traffic clients) first participates in the \textit{registration subprotocol}, which binds available message channels within the cohort with the specific broadcaster. Upon successful registration, only the broadcaster who registered a channel can publish messages to it, until the registration expires. {\pepper} employs cryptographic mechanisms to ensure that broadcasters can anonymously verify the success of their registrations and prevent unauthorized writes to registered message channels.

Following successful registration, the broadcaster transitions to the second sub-protocol, the \textit{messaging subprotocol}. Similar to the registration phase, this phase operates in epochs. During each epoch, the broadcaster anonymously publishes messages while cover-traffic clients generate dummy traffic to preserve anonymity guarantees. The registration phase is a prerequisite for the messaging phase, as it establishes the cryptographic bindings and access control necessary for disruption-resistant anonymous communication.

Next, we provide a step-by-step overview of these two subprotocols, leaving their design details presented later in Section~\ref{sec:system}.

\subsection{Overview of the Registration Subprotocol}
\label{subsec:regi_overview}

The registration subprotocol enables a broadcaster to anonymously acquire one or more messaging channels without revealing their network identity. This process cryptographically binds the broadcaster's public key to specific message channels, which are later used to enforce write permissions during messaging. The protocol consists of two main phases: a \textit{setup phase} and a \textit{registration phase}.

\subject{Setup Phase.}  
Prior to registration, the Pepper system publishes a list of available server pairs (called \textit{cohorts}) through a trusted directory service. Each cohort consists of two aggregation servers that will jointly process registration requests. Clients query the directory to retrieve a cohort and corresponding network paths to each server. A lightweight connectivity check (\textit{dialing}) ensures that both servers in the selected cohort are reachable and ready to accept requests.

\textbf{Registration Phase.}  
Once a cohort is selected, the broadcaster constructs a \textit{registration message} containing:
\begin{itemize}
    \item A public key $g^a$ (the broadcaster’s long-term identity),
    \item A starting message channel identifier $c$ and requested bandwidth $b$ (number of channels),
    \item A cryptographic hash $H(g^a, c, b)$ to ensure integrity.
\end{itemize}
This message is \textit{secret-shared} between the two aggregation servers using a \textit{verifiable distributed point function (VDPF)}~\cite{boyle2016function, de2022lightweight}—a cryptographic primitive that allows compact, verifiable secret sharing. The broadcaster sends one share to each server via separate network paths.

Upon receiving shares, the two servers perform a \textit{joint audit} to verify that the shares are well-formed and correspond to a valid registration request. This audit is lightweight: servers exchange small verification tokens and check consistency without learning the content of the request. Only requests that pass the audit are retained.

Finally, each server aggregates all valid shares it has received. After aggregation, the servers exchange their aggregated results and reconstruct the full set of registration messages. Each valid registration is then published to a public bulletin, binding the broadcaster’s public key $g^a$ to the requested channel(s). Once published, the broadcaster can verify whether their registration succeeded or not by checking the bulletin, while remaining anonymous.

The entire process preserves sender anonymity: the two aggregation servers see only secret shares, and even a global adversary observing all network traffic cannot link a registration request to the originating broadcaster.

\subject{Design Challenges and Cryptographic Solutions.}
Several challenges arise in designing such a registration protocol: (i) \textit{Registration Conflicts:} How to prevent multiple clients from concurrently registering the same channel identifier, and how to resolve contention fairly and anonymously?
(ii) \textit{Joint Auditing:} How can two non-colluding servers efficiently verify the validity of secret shares without reconstructing the original message or learning its target channel?

We address these challenges with carefully designed cryptographic mechanisms, including the use of verifiable secret sharing, collision-resistant hashing, and a structured message format that supports unambiguous channel assignment. The technical details of these mechanisms are presented in Section~\ref{subsec:registration}.

\subsection{Overview of the Messaging Subprotocol}
\label{subsec:msg_overview}
Following successful registration, a broadcaster can anonymously publish messages to their registered channels. The messaging subprotocol operates in three sequential phases: \textit{message sharing}, \textit{auditing}, and \textit{recovery}. 

\subject{Message Sharing Phase.}
To send a message, the broadcaster prepares a \textit{message package} that includes: (1) A \textit{verifiable secret share} of the message, encoded so that only the intended channel(s) can later reconstruct it; (2) A cryptographic proof that demonstrates registration of the target channel(s) without revealing the private key.

Cover-traffic clients follow a similar process but send dummy messages to designated \textit{phantom channels} (Section~\ref{subsec:messaging})—special channels that do not correspond to real bulletin entries and serve only to generate cover traffic. 

\subject{Auditing Phase.}
Upon receiving a message share, each server performs a lightweight \textit{joint audit} with the other server to verify two properties without reconstructing the message:
\begin{enumerate}
    \item \textit{Format correctness:} The share is well-structured and matches the expected cryptographic format.
    \item \textit{Authorization:} The client possesses the right to write to the claimed channel(s).
\end{enumerate}
This audit is conducted using compact verification tokens exchanged between the servers. Shares that fail either check are discarded, ensuring that only valid and authorized writes proceed to the next phase. The auditing process preserves privacy: neither server learns the message content; nor do they learn the relationship between a broadcaster and her registered channels.

\subject{Recovery Phase.}
Valid shares are accumulated by each server into a per-channel aggregate. Upon auditing, the two servers exchange their aggregated results and collaboratively reconstruct the final messages for each channel. These messages are then published to the public bulletin, where any subscriber can read themx.

\subject{Support for Batch Messaging.}
A key feature of Pepper is \textit{batch messaging}: a single request can carry multiple messages (up to a parameter \(t\)) to different channels registered under the same broadcaster. This is enabled by a \textit{multi-channel encoding mechanism} that generalizes the single-channel secret-sharing approach. Batching reduces per-message auditing overhead and significantly improves throughput, making Pepper suitable for high-bandwidth applications such as file sharing or media dissemination.

\subject{Design Challenges and Cryptographic Approach.}
Several challenges arise in building a practical messaging subprotocol:
\begin{itemize}
    \item \textit{Access Control:} How to ensure only the owner of a channel can write to it, while still allowing cover traffic?
    \item \textit{Audit Efficiency:} How to verify message legitimacy without expensive per-message cryptographic operations?
    \item \textit{Batch Messaging:} How to deliver multiple messages in a single broadcaster write without proportionally increasing auditing cost?
\end{itemize}
We address these through a combination of verifiable secret sharing, zero-knowledge-inspired proof techniques, and an efficient batch-audit mechanism. The detailed cryptographic constructions are presented next in Section~\ref{sec:system}.

\section{The Pepper System}
\label{sec:system}

This section presents a detailed description of the Pepper protocol. We first outline a protocol template (i.e., the fundamental protocol) in Section~\ref{subsec:primary}, which is then customized into the registration variant (Section~\ref{subsec:registration}) and the messaging variant (Section~\ref{subsec:messaging}), leveraging necessary cryptographic primitives to satisfy the design requirements of registration and messaging respectively. 

\subsection{The Fundamental Protocol}
\label{subsec:primary}

Pepper's core mechanism builds on the integration of distributed point functions (DPFs) with DC-net style secret sharing. This foundation enables efficient, verifiable anonymous writes while maintaining strong privacy guarantees. In this section, we first introduce how DPFs can support multiple broadcast channels with minimal bandwidth overhead, then discuss our fundamental protocol template underpinning both the registration and messaging subprotocols.

\subject{Distributed Point Functions.} A Distributed Point Function (DPF)~\cite{gilboa2014distributed} is a special case of Function Secret Sharing (FSS)~\cite{boyle2015function, boyle2016function} where the function family $\mathcal{F}$ consists of point functions, namely functions $f_{\alpha, \beta}$ that evaluate to $\beta$ on input $\alpha$ and to $0$ on all other inputs. An $n$-party DPF scheme splits a point function $f: X \rightarrow \mathbb{F}$ for a finite field  $\mathbb{F}$ into functions $f_1, f_2, \dots, f_n$ described by keys $k_1, k_2, \dots, k_n$, where $X = \{0,1\}^\ell$ for some $\ell \in \mathbb{N}$. The original point value $f(\alpha)$ can be recovered by additively combining $f_1(\alpha), f_2(\alpha), \dots, f_n(\alpha)$, while every strict subset of the keys reveals no information about $f$.

\begin{definition}[Distributed Point Functions (DPFs)]
    Let $L$ be the domain size of the $n$-party point function, $\lambda$ be the security parameter, and $\mathcal{B}$ be the message space (e.g., a finite field of prime order $\mathbb{F}$). The function scheme consists of two PPT algorithms for all $i \in \{1, 2, ..., n\}$:
    \begin{itemize}
        \item $\mathsf{Gen}(1^\lambda, \alpha \in \{1, 2, ..., L\},\beta \in \mathcal{B}) \rightarrow (k_1, k_2, ..., k_n)$
        \item $\mathsf{Eval}(k_i) \rightarrow (\beta_{i,1}, \beta_{i,2}, ..., \beta_{i,L})$ 
    \end{itemize}

    These algorithms must satisfy the following properties:
    \begin{itemize}
        \item \textbf{Correctness.} \textit{A correct DPF must ensure that the sum of evaluation outputs equals $\beta$ at point $\alpha$ and $0$ for all other inputs}: 
        $$\Pr\left[
        \begin{array}{cc}
            (k_1, k_2, ..., k_n) \leftarrow \mathsf{Gen}(1^\lambda, \alpha, \beta)  \\
            (\beta_{i,1}, \beta_{i,2}, ..., \beta_{i,L}) \leftarrow \mathsf{Eval}(k_i) \\
            \sum_{i=1}^n \beta_{i,j} = 0,\; j\neq \alpha  \\
            \sum_{i=1}^n \beta_{i,j} = \beta,\; j = \alpha
        \end{array} \right] = 1$$
        \item \textbf{Privacy.} \textit{A DPF is private if any strict subset of evaluation keys reveals no information about the inputs, meaning there exists an efficient simulator $\mathsf{Sim}$ that generates outputs computationally indistinguishable from strict subsets of keys output by $\mathsf{Gen}$}.

        \item \textbf{Compactness.}
        \textit{The key size must be sublinear in the domain size (i.e., the number of channels in broadcast).} 
    \end{itemize}
\end{definition}

In Pepper, DPFs serve as the underlying mechanism for anonymous writes. Instead of sending raw message shares as in classic DC-nets, clients use DPFs to encode their messages at specific channel positions. This approach provides several key advantages: (1) \textit{Bandwidth efficiency}: clients only send short DPF keys rather than full message vectors, significantly reducing communication overhead; (2) \textit{Verifiability}: servers can verify that DPF shares are well-formed before aggregation, enabling efficient filtering of malicious requests; (3) \textit{Channel privacy}: servers learn which channels are being written to only after aggregation, not from individual shares, preserving sender anonymity during transmission.

\begin{figure}
    \centering
    \includegraphics[width=\linewidth]{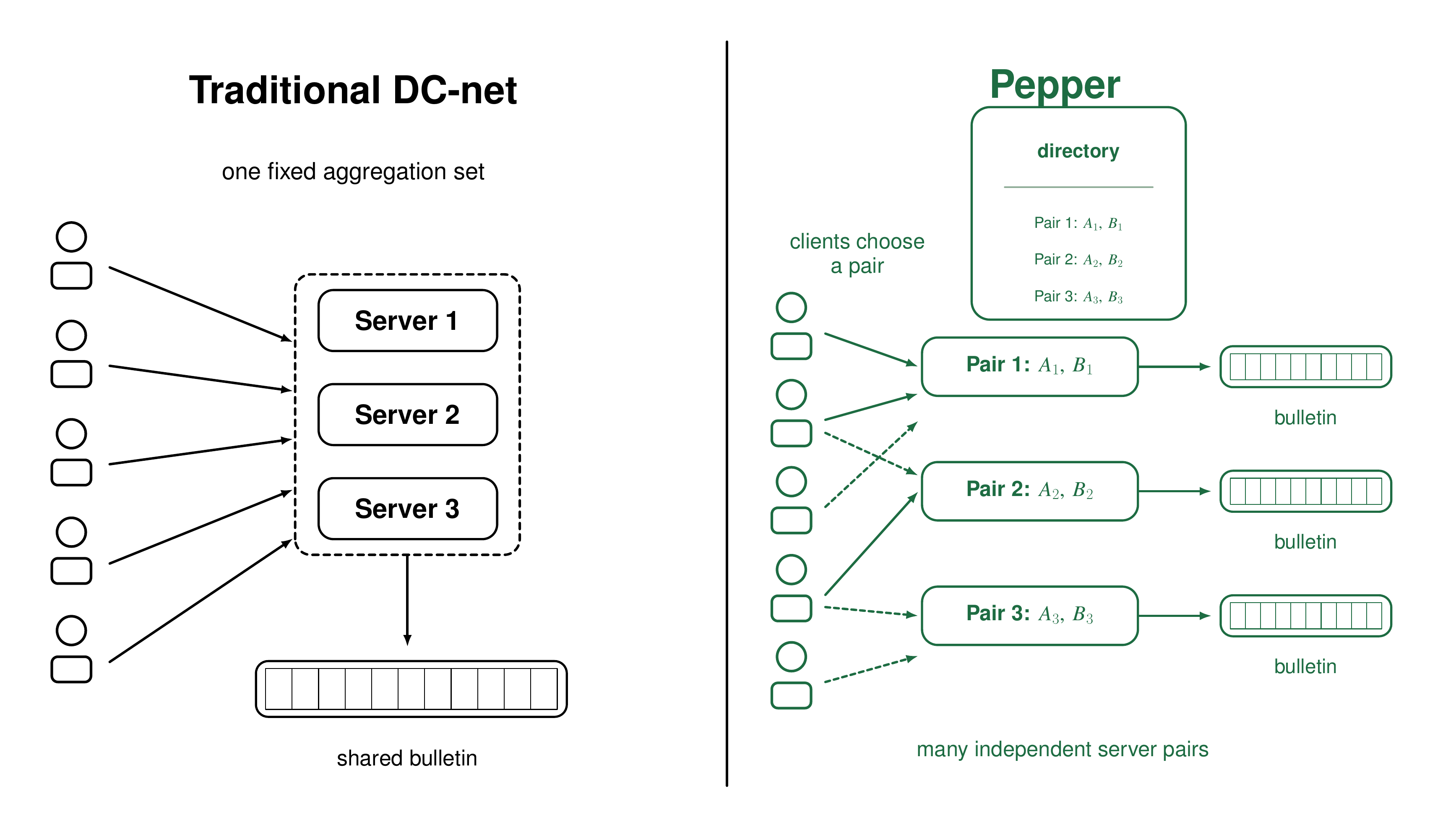}
    \caption{Pepper's deployment architecture compared with a traditional server-based DC-net. Traditional DC-nets route clients through one fixed aggregation set that publishes to a shared bulletin. Pepper instead exposes multiple independent aggregation server pairs through a directory service; clients select one pair, and each pair audits and aggregates requests for its own bulletin.}
    \label{fig:pepper_architecture}
\end{figure}

\subject{Protocol Flow.} The fundamental protocol template operates in three phases: (1) \textit{Share generation}: clients create DPF shares for their messages and target channels; (2) \textit{Auditing}: servers verify share validity and authorization (this phase is enhanced with verifiability in subsequent protocols); (3) \textit{Aggregation}: servers combine valid shares to reconstruct messages. This foundation supports both registration (binding public keys to channels) and messaging (publishing to registered channels) subprotocols. Pepper's efficiency stems from the compact message share representation and parallelizable evaluation, enabling high-throughput anonymous communication that scales better than traditional DC-net approaches. Figure~\ref{fig:pepper_architecture} summarizes the deployment-level distinction between Pepper and a traditional DC-net: Pepper clients select among multiple independent aggregation server pairs rather than relying on a single fixed aggregation set.

\subsection{The Registration Subprotocol}
\label{subsec:registration}

The registration protocol enables broadcasters to anonymously establish their identity within the Pepper system by binding their public keys to specific messaging channels. However, in a malicious setting, clients may attempt to send nonsense or malformed messages to disrupt the system. To prevent such attacks, we need a mechanism that allows servers to verify that message shares are well-formed before aggregation, without learning which specific message channel is being targeted. This requirement leads us to Verifiable Distributed Point Functions (VDPFs)~\cite{boyle2016function, de2022lightweight}. Next, we first provide an overview of VDPF before elaborating the details of the registration subprotocol.



\subject{Verifiable Distributed Point Functions.} Verifiable DPFs (VDPFs)~\cite{boyle2016function, de2022lightweight} extend standard DPFs by incorporating a hash digest of the DPF key evaluation path during key generation. This digest allows servers to jointly verify that each DPF key encodes a well-formed point function with exactly one valid point, ensuring that all verifiers hold shares corresponding to the same point, without revealing the target point. Specifically, during evaluation, each server expands the DPF tree and computes hash values based on its hash digest. If the digest is correctly generated and all shares originate from the same VDPF key generation, the final hash values computed by all servers will match. Since the hash digest and the verification token $\rho_i$ output by $\mathsf{Eval}$ are of fixed size depends on the security parameter $\lambda$, their overhead is negligible. The VDPF scheme is defined as follows:

\begin{definition}[Verifiable Distributed Point Functions (VDPFs)]
     Let $L$ be the domain size of the $n$-party point function, $\lambda$ be the security parameter, and $\mathcal{B}$ be the message space. The function scheme consists of three PPT algorithms for all $i \in \{1, 2, ..., n\}$:
    \begin{itemize}
        \item $\mathsf{Gen}(1^\lambda, \alpha \in \{1, 2, ..., L\},\beta \in \mathcal{B}) \rightarrow (k_1, k_2, ..., k_n)$
        \item $\mathsf{Eval}(k_i) \rightarrow (\beta_{i,1}, \beta_{i,2}, ..., \beta_{i,L}, \rho_i)$ 
        \item $\mathsf{Verify}(\rho_1, \rho_2, \dots, \rho_n) \rightarrow 1/0$
    \end{itemize}
    These algorithms must satisfy the following properties if the VDPF is secure:
    \begin{itemize}
        \item \textbf{Correctness.} \textit{A correct VDPF must ensure that the sum of evaluation outputs equals $\beta$ at point $\alpha$ and $0$ for all other inputs, and that verification succeeds}: 
        $$\Pr\left[
        \begin{array}{cc}
            (k_1, k_2, ..., k_n) \leftarrow \mathsf{Gen}(1^\lambda, \alpha, \beta)  \\
            (\beta_{i,1}, \beta_{i,2}, ..., \beta_{i,L}, \rho_i) \leftarrow \mathsf{Eval}(k_i) \\
            \sum_{i=1}^n \beta_{i,j} = 0,\; j\neq \alpha  \\
            \sum_{i=1}^n \beta_{i,j} = \beta,\; j = \alpha \\
            \mathsf{Verify}(\rho_1, \rho_2, \dots, \rho_n) =1
        \end{array} \right] = 1$$
        \item \textbf{Privacy.} \textit{A VDPF is private if any strict subset of evaluation keys reveals no information about the inputs, meaning there exists an efficient simulator $\mathsf{Sim}$ that generates outputs computationally indistinguishable from strict subsets of keys output by $\mathsf{Gen}$}.

        \item \textbf{Soundness.} \textit{Soundness guarantees that if verification passes, the evaluated function is well-formed, meaning it has at most one valid point. Formally, for any set of keys, the probability that verification accepts an invalid function (with more than one non-zero position) is negligible:}
        $$\Pr\left[
        \begin{array}{cc}
        \forall i: (\vec{\beta}_i, \rho_i) \leftarrow \mathsf{Eval}(k_i) \\
        \mathsf{Verify}(\rho_1, \dots, \rho_n) = 1 \\
        \left|\left\{j \in \{1, \dots, L\} : \sum_{i=1}^{n} \beta_{i,j} \neq 0\right\}\right| \leq 1
        \end{array} \right] \geq 1 - \mathsf{negl}(\lambda)$$

        \textit{This means that with overwhelming probability, if \(\mathsf{Verify}\) outputs 1, at most one position has a non-zero sum of the evaluations.}

    \end{itemize}
\end{definition}

The Pepper system employs two types of channels:

\begin{itemize}
    \item \textit{Registration Channels.} Used by broadcasters to anonymously register messaging channels and bind public keys to selected channels. We denote the number of registration channels as $L_1$. A registration is considered successful when servers complete verification and successfully recover the registration message, confirming that the broadcaster's public key has been published to the system.
    \item \textit{Message Channels.} Used for actual message publishing after successful registration. We denote the number of message channels as $L_2$, where $L_2 \gg L_1$ to support high-bandwidth communication and to make registration collisions on any individual message channel unlikely in practice.
\end{itemize}



\begin{figure}
    \centering
    \includegraphics[width=\linewidth]{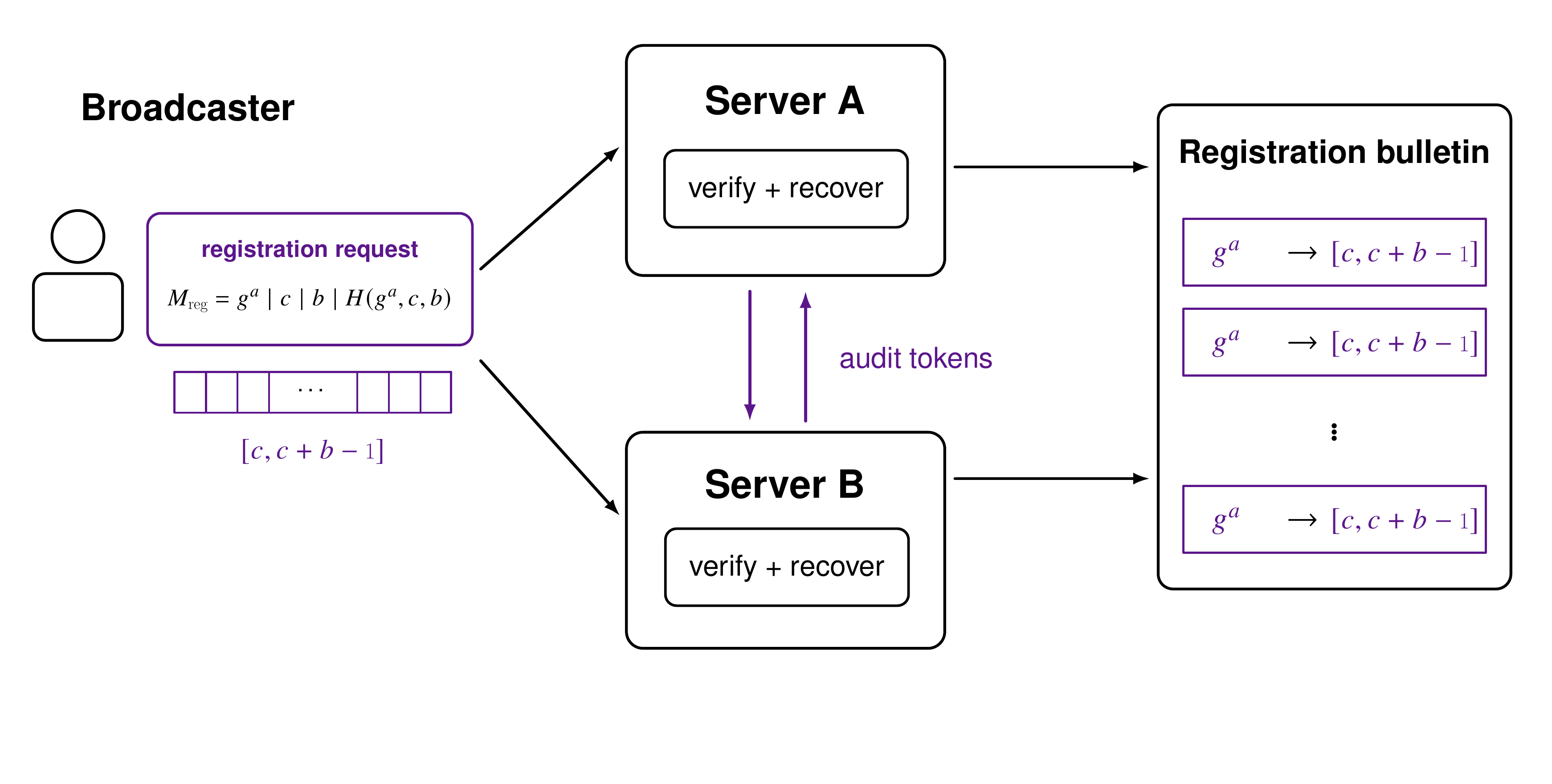}
    \caption{Registration subprotocol. A broadcaster encodes a registration request containing its public key $g^\alpha$, starting message channel $c$, requested bandwidth $b$, and integrity hash. The two aggregation servers exchange audit tokens, verify and recover valid registrations, and publish bindings from $g^\alpha$ to the registered channel range $[c,c+b-1]$ on the registration bulletin.}
    \label{fig:pepper_registration_flow}
\end{figure}

The registration message format is designed to bind the broadcaster's public key $g^\alpha$ to specific messaging channels while enabling efficient verification. To construct a registration message $m$, the broadcaster first selects a starting message channel identifier $c_{l_2}$ and specifies the requested bandwidth $b$, which represents the number of consecutive channels to register starting from $c_{l_2}$. This enables batch registration of a range of channels $[c_{l_2}, c_{l_2}+b]$ in a single request. The broadcaster then includes their public key $g^\alpha$, which establishes their cryptographic identity and will be used for access control during messaging. Finally, the broadcaster computes a collision-resistant hash $\mathcal{H}$ over all components to ensure message integrity and prevent tampering during transmission. The complete registration message is assembled as:
$$m \leftarrow g^\alpha \| c_{l_2} \| b \| \mathcal{H}(g^\alpha \| c_{l_2} \| b)$$
This design enables servers to verify that the registration request is authentic and well-formed without learning the broadcaster's identity or target messaging channels during the auditing phase. Figure~\ref{fig:pepper_registration_flow} illustrates how the two servers audit, recover, and publish the resulting public-key-to-channel-range binding.

The registration protocol operates through the following steps:

\textbf{Step 0-1: Server Pair Publication.} A trusted directory service publishes information about available pairs of aggregation servers. Given servers supporting the protocol (e.g., existing Tor relays), the directory service can randomly designates which two servers compose an aggregation pair to avoid collusion. For each pair, the service publishes server addresses, connection credentials (e.g., TLS public keys), and periodically updates the number of connecting clients, which helps broadcasters select pairs with a large anonymity set.

\textbf{Step 0-2: Dialing.} Clients query the directory service to retrieve available aggregation server pairs and select a target pair. A connectivity test (ping) is sent through both paths; if successful, the servers are confirmed active and registration can proceed. We denote the two aggregation servers in the selected pair as Server 1 and Server 2. This dialing mechanism ensures that clients can dynamically adapt to server availability and network conditions, improving system resilience compared to fixed server assignments.

\textbf{Step 1: Message Sharing.} A broadcaster selects a registration channel $l_1 \in \{1, 2, \ldots, L_1\}$ and a message channel $l_2 \in \{1, 2, \ldots, L_2\}$ denoting the starting message channel id and bandwidth $b$ denoting the number of message channels to register. As the message channel space ($L_2$) is very large and the bandwidth is far smaller than $L_2$, the collision probability tends to be negligible for such a batching registration process. Once the registration succeeds, message channels $[l_2, l_2 + 1, l_2 + 2, ..., l_2 + b - 1]$ will be registered by the client along with these message channels bound with the client's public key. 

All clients share a common collision-resistant hash function $\mathcal{H}: \{0,1\}^* \rightarrow \{0,1\}^\lambda$ for message integrity verification. The client constructs the registration message and generates VDPF shares:
\begin{enumerate}
    \item $m \leftarrow g^\alpha \| l_2 \| b \| \mathcal{H}(g^\alpha \| l_2 \| b)$
    \item $(k_1, k_2) \leftarrow \mathrm{VDPF.Gen}(1^\lambda, m, l_1)$
\end{enumerate}

\textbf{Step 2: Message Auditing.} Upon receiving a message share from a client, each server $i$ performs verification:
\begin{enumerate}
    \item $(\vec{m}_i, \rho_i) \leftarrow \mathrm{VDPF.Eval}(k_i)$ where $\vec{m}_i$ denotes the vector of $L_1$ message shares expanded from the key material for server $i$.
    \item Exchange $\rho_i$ with the other server to complete verification
    \item If $\mathrm{VDPF.Verify}(\rho_1, \rho_2) = 0$, drop the message
\end{enumerate}
Pepper's verification process requires a small-size token exchange between servers, making it highly efficient compared to proof-based systems like Express that require large proof transfer. This lightweight verification is a key advantage of Pepper's registration protocol.

\textbf{Step 3: Message Recovery.} Each server maintains an accumulator $acc_i$ of $L_1$ entries, each corresponding to a registration channel, to accumulate all valid requests that pass Step 2 auditing. The accumulator is defined as $acc_i = \sum_{j \in \mathcal{C}} \vec{m}_{i,j}$, where $\mathcal{C}$ is the set of all clients whose requests passed auditing, and $\vec{m}_{i,j}$ is the vector of $L_1$ message shares from client $j$ for server $i$. Here, the summation is carried out channel-wise, i.e., $acc_i[k] = \sum_{j \in \mathcal{C}} \vec{m}_{i,j}[k]$ with $k$ defined as the registration channel id. Then, to recover the registration messages:
\begin{enumerate}
    \item Publicly reveal and exchange $acc_i$
    \item After receiving all message shares from the other server, recover the vector of $L_1$ messages via $\vec{m} = acc_1 \oplus acc_2$
\end{enumerate}

After reconstruction, servers parse each recovered registration record $m = g^\alpha \| l_2 \| b \| h$ and independently recompute $\mathcal{H}(g^\alpha \| l_2 \| b)$ to check whether it equals $h$. If this check succeeds and the message channel range $[l_2, l_2 + b - 1]$ has yet to be registered, the servers bind them with $g^\alpha$. The resulting registration records are then published on the bulletin for each registration channel, so that clients can confirm registration success. If two or more broadcasters collide on the same registration slot, or if the hash verification fails, the servers discard the corresponding records for the current epoch. In this case, the broadcaster must wait for the next registration round to select fresh registration channels and retry until her registration is accepted.

\subsection{The Messaging Subprotocol}
\label{subsec:messaging}

The messaging protocol enables registered broadcasters to anonymously publish messages while attaining anonymity with the assistance from cover-traffic clients, e.g., subscribers in a microblogging platform, or volunteering Tor users. 

However, a naive server-based DC-net design is vulnerable to write disruptions, as any client including malicious ones can write a non-zero message to a channel and thus ``overwrite'' the intended message from the broadcaster, rendering the necessity of access control enforcement. Besides, although recent server-based DC-nets (e.g., Spectrum and Express) support messaging from multiple broadcasters per epoch, each broadcaster is only allowed to write a single message to a single channel in each epoch, impeding \textit{message batching}. Below, we elaborate the messaging subprotocol of {\pepper} with an emphasis on its lightweight access control and support for  message batching.  


\subject{Setup.} After the registration phase, the system exposes $L$ registered message channels, indexed by a verification-key vector $(g^{\alpha_1}, g^{\alpha_2}, \ldots, g^{\alpha_L})$ shared among all servers. Typically, $L$ is subject to $L \ll L_2$ where $L_2$ is aforementioned message channel space. For clarity, we describe the messaging subprotocol for a two-server setup and focus on the scenario where a single broadcaster  wishes to send a message $m$ to one or more target channels associated with a specific public key.



Next, we progressively introduce our messaging subprotocol with new features added gradually. We first present the initial version as inspired by Spectrum~\cite{newman2022spectrum}, and we name it as \textit{Messaging V1}. We then present \textit{Messaging V2}, which further optimizes the audit efficiency upon \textit{Messaging V1}. Finally, we revise \textit{Messaging V2} with support of batch messaging, leading to our final messaging subprotocol \textit{Messaging V3}.       


\subsubsection{Messaging V1}
\label{subsub:spectrum}

As outlined in Section~\ref{subsec:msg_overview}, anonymous messaging requires servers to perform joint auditing that verifies two properties: (i) \textit{Format correctness}:  Ensuring that DPF shares are well-formed and match the expected cryptographic format;  (ii) \textit{Authorization}: Ensuring that clients possess the right to write to the claimed channels. Prior work typically addressed these two requirements separately~\cite{corrigan2015riposte, eskandarian2021express}. Spectrum~\cite{newman2022spectrum} was the first protocol to combine both verifications in a unified access control mechanism, enabling servers to efficiently verify both the correctness of DPF shares and the authorization of clients to write to specific channels in a single auditing step. Our draft messaging subprotocol starts with that from Spectrum. Next, we review Spectrum as the foundation for our final messaging subprotocol.


Notably, this initial messaging subprotocol only supports broadcasting a single message to a single channel. The main flow consists of the following steps:

\textbf{Step 1: Message Sharing.} Given a broadcaster and her private key $\alpha_j$, registered channel id $j$ and a non-zero message $m$ to broadcast. Let $y = \alpha_j$ and $j' = j$ for this broadcaster while setting up $y = 0$ and $j' = 0$ for cover-traffic clients. Only broadcasters have non-zero messages $m \neq 0$. Each client performs the following operations:
\begin{enumerate}
    \item $(k_1, k_2) \leftarrow \mathrm{DPF.Gen}(1^\lambda, m, j')$
    \item $t \leftarrow m \cdot y$
    \item Sample $\text{linear shares } (t_1, t_2) \leftarrow t$
    \item $\text{Send } (k_i, t_i) \text{ to } \text{server }i \text{ for } i \in \{1,2\}$
\end{enumerate}

\textbf{Step 2: Message Auditing.} Once servers receive a request share pair $(k_i, t_i)$ from a client, each server runs:
\begin{enumerate}
    \item $m_{i} \leftarrow \mathrm{DPF.Eval}(k_i)$
    \item $A_i \leftarrow \prod_{j=1}^L {g^{ m_i[j] \cdot {\alpha_j} }}$
    \item $B_i \leftarrow A_i/g^{t_i}$
    \item Send $B_i$ to another server
\end{enumerate}
All servers check if $B_1 \cdot B_2 = g^0 = 1$. If the check fails, servers drop this request share. 

\textbf{Step 3: Message Recovery.} Each server maintains an accumulator $acc_i$ of $L$ entries (channels). The accumulator is defined as $acc_i = \sum_{j \in \mathcal{C}} \mathrm{DPF.Eval}(k_j)$, where $\mathcal{C}$ is the set of all clients whose shares passed auditing, and $k_j$ is the DPF key from client $j$:
\begin{enumerate}
    \item Compute $acc_i \leftarrow \sum_{j \in \mathcal{C}} \mathrm{DPF.Eval}(k_j)$
    \item Publicly reveal $acc_i$
\end{enumerate}
Using the publicly revealed shares, anyone can recover the $L$ broadcast messages $\vec{m} = \sum acc_i$.

\subject{Limitations.} On one side, the  audit requires each server to perform $L$ exponentiations and multiplications due to the group operations involved, which becomes computationally expensive as the number of channels increases. Besides, message batching is not supported, as a broadcaster can only submit one message per epoch.

\subsubsection{Messaging V2: Optimizing Audit Efficiency}
\label{subsub:pacl}


 To improve audit efficiency, we further integrates into our messaging subprotocol an efficient access control technique, namely Schnorr Proof over Secret Shares (SPoSS), which was first adopted in VDPF-PACL~\cite{servan2023private}) for enabling access control for function secrete sharing. The integration is non-trivial, as the original mechanism doesn't support cover-traffic clients that have no private key registered. Next, we first provide an overview of this new auditing technique, and then present the revised messaging subprotocol with audit support of cover-traffic clients.



\subject{Schnorr Proof over Secret Shares (SPoSS).} SPoSS~\cite{servan2023private} is a non-interactive zero-knowledge proof-of-knowledge protocol for discrete logarithms over an additively secret-shared group element~\cite{boneh2019zero, corrigan2017prio}.
Let $\mathbb{G}$ be a cyclic group of prime order $q$ with generator $g$. Consider a private key $x \in \mathbb{Z}_q$ and its corresponding public key $y = g^x \in \mathbb{G}$.
In the 2-party setting, $y$ is additively secret-shared between verifiers such that $y = y_1 + y_2$ (where $y_1, y_2 \in \mathbb{G}$).
The protocol comprises three PPT algorithms:
\begin{itemize}
    \item $\mathsf{Prove}(x) \rightarrow (\pi_1, \pi_2)$: Takes the private key $x$ as input and generates proof shares $(\pi_1, \pi_2)$ for the two verifiers.
    \item $\mathsf{Audit}(y_i, \pi_i) \rightarrow \tau_i$: Takes a verifier's key share $y_i$ and the corresponding proof share $\pi_i$ as input, outputting an audit token $\tau_i$.
    \item $\mathsf{Verify}(\tau_1, \tau_2) \rightarrow \{0, 1\}$: Aggregates the audit tokens from both verifiers to validate the proof. It outputs $1$ if and only if the reconstructed values satisfy the underlying discrete logarithm relation (i.e., verifying $y_1 + y_2 = g^x$ implicitly).
\end{itemize}

We treat SPoSS as a black-box primitive, deferring concrete instantiation and security definitions to Appendix~\ref{appendix:sposs}. Given the messaging protocol V1, we integrate SPoSS as follows. Let $\alpha_j$ be private key for channel $j$. For a broadcaster writing to channel $j$ with a non-zero message $m$, the client runs $\mathsf{SPoSS.Prove}(\alpha_j)$ to generate proof shares $(\pi_1, \pi_2)$. The client sends $(k_i, \pi_i)$ to each server $i$, where $k_i$ is the DPF key share. In the auditing phase, server $i$ first evaluates the DPF key to obtain the vector share $\vec{m}_i = \mathrm{DPF.Eval}(k_i)$. The server then computes an accumulator $A_i = \sum_{j=1}^L \vec{m}_i[j] \cdot g^{\alpha_j}$. For a specific channel $j'$, $A_i$ could be considered as an additive share of $g^{{\alpha_j}'}$ when $\vec{m}[j] = 0$ for $j \neq j'$. The server then runs $\tau_i = \mathsf{SPoSS.Audit}(A_i, \pi_i)$ to generate an audit token. The servers exchange $\tau_i$ and verify $\mathsf{SPoSS.Verify}(\tau_1, \tau_2) = 1$. This verification confirms that the broadcaster knows the discrete logarithm $\alpha_j$ corresponding to the target channel, without revealing which channel was targeted.


While SPoSS effectively verifies key ownership, a vulnerability remains regarding the binding between the message and the specific channel identity. Consider a malicious broadcaster with a valid private key $\alpha_1$ who intends to disrupt a victim channel $\alpha_2$. The attacker could craft a message $m$ and proof shares that algebraically satisfy the SPoSS relation (e.g., by manipulating $m \approx g^{\alpha_1}/g^{\alpha_2}$), potentially tricking the audit into accepting a message that disrupts the target. To eliminate such algebraic manipulation, we introduce a $1$-bit checksum. We mandate that the first bit of the message payload must match a specific derivation from the key. This binds the payload value strictly to the channel identity and any attempt to "shift" the target channel via algebraic division would invalidate the checksum, causing the subsequent verification to fail.

Furthermore, integration of SPoSS requires that every participating client knows a private key bound to a message channel, which, however, is not the case for cover-traffic clients, impeding cover-traffic clients to generate qualified client requests.  To address this problem, We introduce \textit{Phantom Channels} to enable qualified cover traffic generation. We observe that for a trivial public key (e.g., $\{g^1, g^2, \dots\}$), the discrete logarithm is universally known. During system setup, we reserve a set of Phantom Channels where the implied private key is set to $\{1, 2, \dots\}$, at the end of registration or beginning of messaging, the Pepper system generates a list of such channels for cover traffic that everyone can use to send messages that will never appear in bulletins. For cover traffic, clients set their target to a Phantom Channel and compute the proof term for any message payload. This allows the cover traffic passes SPoSS's audit without requiring a registered user identity.


The main flow of our \textit{Messaging V2} is as follows:

\textbf{Step 1: Message Sharing.} Let $y = \alpha_j$ and $j' = j$. Only broadcasters have non-zero messages $m \neq 0$. Each client performs the following operations:
\begin{enumerate}
    \item $m' \leftarrow 1^{\text{bit}} \| m$ 
    \item $(k_1, k_2) \leftarrow \mathrm{VDPF.Gen}(1^\lambda, m', j')$ 
    \item $(\pi_1, \pi_2) \leftarrow \mathrm{SPoSS.Prove}(y)$
    \item Encrypt share-pair $(i, k_i)$ and send to server $i$ via onion routing path, for $i \in \{1, 2\}$
\end{enumerate}

\textbf{Step 2: Message Auditing.} After all nodes in the path correctly decrypt the message and server $i$ receives share $(i, k_i)$ from a client, each server performs:
\begin{enumerate}
    \item $(m_{i}', \rho_i) \leftarrow \mathrm{VDPF.Eval}(k_i)$
    \item $m_i^\text{bit} \leftarrow \text{Aggregation of first bit of } m'_i[j],\; j \in \{1, \dots, L\}$
    \item $A_i \leftarrow \sum_{j=1}^L m_i^{bit}[j] \cdot g^{\alpha_j}$
    \item $\tau_i \leftarrow \mathrm{SPoSS.Audit}(A, \pi_i)$ 
    \item Send $(\rho_i, \tau_i)$ to the other server
\end{enumerate}
All servers verify: $\mathrm{SPoSS.Verify}(\tau_1, \tau_2) = 1$, and $\mathrm{VDPF.Verify}(\rho_1, \rho_2) = 1$. If verification passes, proceed to Step 3.

\textbf{Step 3: Message Recovery.} Each server maintains an accumulator $acc_i$ of $L$ entries (channels). The accumulator is defined as $acc_i = \sum_{j \in \mathcal{C}} \mathrm{VDPF.Eval}(k_j)$, where $\mathcal{C}$ is the set of all clients whose shares passed auditing, and $k_j$ is the VDPF key from client $j$:
\begin{enumerate}
    \item Compute $acc_i \leftarrow \sum_{j \in \mathcal{C}} \mathrm{VDPF.Eval}(k_j)$
    \item Publicly reveal $acc_i$
\end{enumerate}
Using the publicly revealed shares, anyone can recover the $L$ broadcast messages $m = \sum acc_i$.


\subsubsection{Messaging V3: Supporting Batch Messaging}

\begin{figure}
    \centering
    \includegraphics[width=\linewidth]{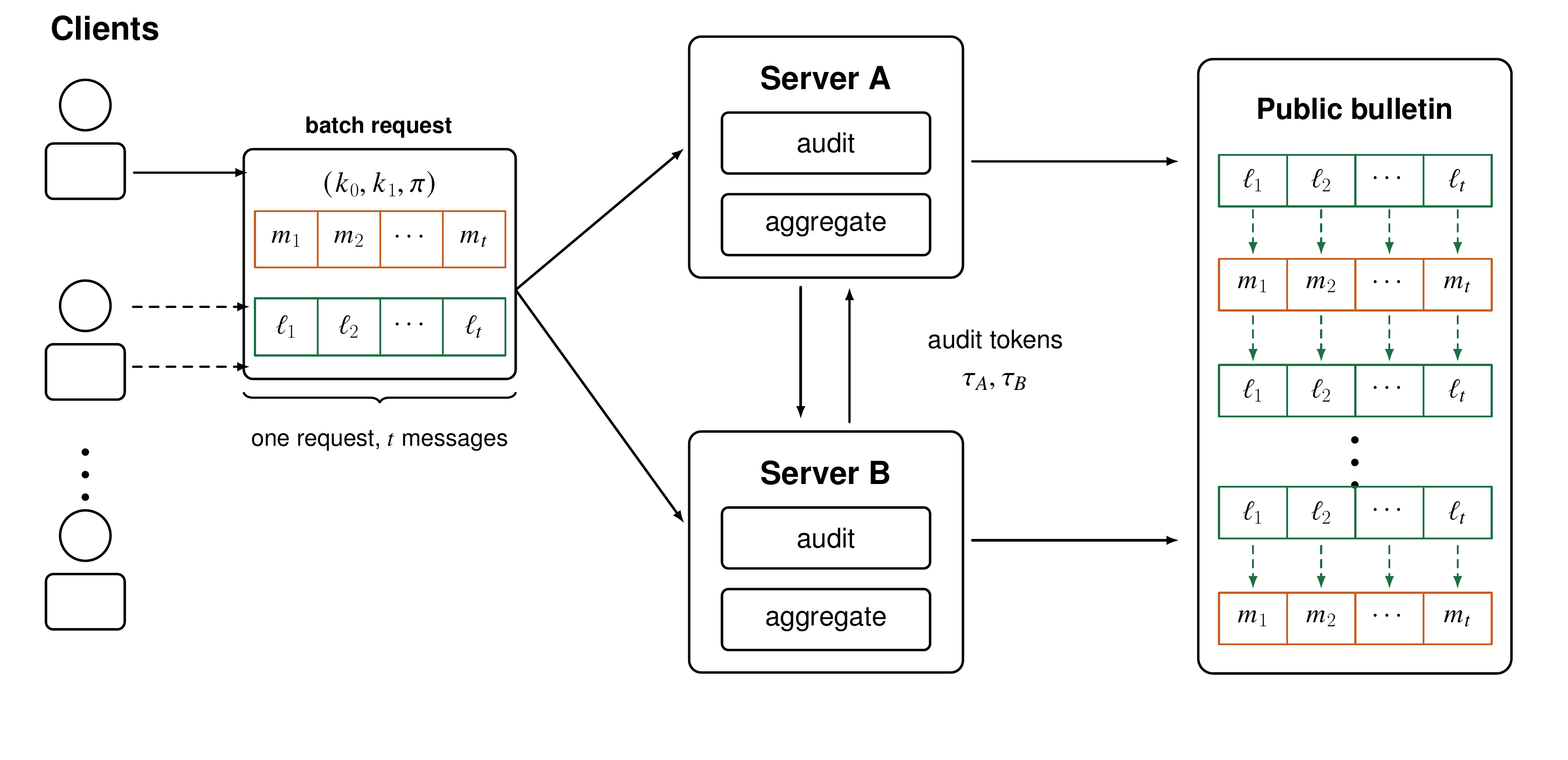}
    \caption{Batch messaging subprotocol. A client sends one batch request containing compact VDMPF key shares, one proof, and a payload vector $(m_1,m_2,\ldots,m_t)$ for target channels $(\ell_1,\ell_2,\ldots,\ell_t)$. The two servers exchange audit tokens once for the batch, aggregate valid requests, and publish multiple channel outputs on the public bulletin.}
    \label{fig:pepper_messaging_batch_flow}
\end{figure}

To support batch messaging, a straightforward approach is to concatenate all messages into a single vector. To preserve anonymity, these vectors must be of uniform length, effectively bounded by the maximum registered bandwidth among all broadcasters. However, this incurs significant communication overhead due to the padding required for smaller requests. Alternatively, Servan, et al.~\cite{servan2023private} introduced VDMPF-PACLs, which considers a Verifiable Distributed Multiple-point Functions (VDMPFs) key as a naive aggregation of multiple VDPF keys. Consequently, to send $t$ messages, the broadcaster must generate $t$ separate proofs, and servers are forced to execute the auditing protocol $t$ times. In contrast, Pepper employs a more efficient Distributed Multi-point Functions Key construction~\cite{boyle2025improved}. This design generates a single consolidated proof for the entire batch, significantly reducing the verification burden while enabling batch messaging, leading to our final messaging subprotocol (\textit{Messaging V3}). Figure~\ref{fig:pepper_messaging_batch_flow} shows the resulting batch flow, where one client request carries multiple messages for multiple target channels and the server pair audits the batch once before publishing the outputs.

Next, we first provide the formal definitions of Distributed Multi-point Functions (DMPF) and Verifiable Distributed Multi-point Functions (VDMPF), and subsequently details how Pepper supports efficient batch messaging.

\subject{Distributed Multi-Point Functions.} Distributed Multi-Point Functions (DMPFs)~\cite{boyle2016function, boyle2025improved} extend DPFs to support evaluation to non-zero values on multiple inputs, enabling efficient multi-channel communication. For a DMPF with $m$ valid inputs $\{\alpha_1, \alpha_2, \dots, \alpha_m\}$ and corresponding outputs $\{\beta_1, \beta_2, \dots, \beta_m\}$, we define the DMPF scheme as follows:

\begin{definition}[Distributed Multi-Point Functions (DMPFs)] 
    \textit{Let $L$ be the domain size of the $n$-party multi-point function, $\lambda$ be the security parameter, and $\mathcal{B}$ be the message space. The function scheme consists of 2 PPT algorithms for all $i \in \{1, 2, ..., n\}$:}
    \begin{itemize}
        \item $\mathsf{Gen}(1^\lambda, \{\alpha_1, \alpha_2, \dots, \alpha_m\} \in \{1, 2, ..., L\},\{\beta_1, \beta_2, \dots, \beta_m\} \in \mathcal{B}) \rightarrow (k_1, k_2, ..., k_n)$
        \item $\mathsf{Eval}(k_i) \rightarrow (\beta_{i,1}, \beta_{i,2}, ..., \beta_{i,L})$ 
    \end{itemize}
    \textit{These algorithms must satisfy the following properties if the DMPF is secure:}
    \begin{itemize}
        \item \textbf{Correctness.} \textit{A correct DMPF must ensure that the sum of evaluation outputs equals $\beta_j$ at point $\alpha_j$ for $j \in \{1, 2, ..., m\}$ and $0$ for all other inputs}: 
        $$\Pr\left[
        \begin{array}{cc}
             (k_1, k_2, ..., k_n) \leftarrow \mathsf{Gen}(1^\lambda, \{\alpha_1, \alpha_2, \dots, \alpha_m\}, \{\beta_1, \beta_2, \dots, \beta_m\})  \\
             (\beta_{i,1}, \beta_{i,2}, ..., \beta_{i,L}) \leftarrow \mathsf{Eval}(k_i) \\
             \sum_{i=1}^n \beta_{i,j} = 0,\; j\notin \{\alpha_1, \alpha_2, \dots, \alpha_m\}  \\
             \sum_{i=1}^n \beta_{i,j} = \beta_k,\; j = \alpha_k \text{ for } k \in \{1, 2, \dots, m\}
        \end{array} \right] = 1$$
        \item \textbf{Privacy.} \textit{A DMPF is private if any strict subset of evaluation keys reveals no information about the inputs, meaning there exists an efficient simulator $\mathsf{Sim}$ that generates outputs computationally indistinguishable from strict subsets of keys output by $\mathsf{Gen}$}.
    \end{itemize}
\end{definition}
\subject{Verifiable Distributed Multi-Point Functions.} DMPFs schemes can be made verifiable like DPFs, researchers have developed various implementation methods for Verifiable DMPFs (VDMPFs)~\cite{boyle2016function, de2022lightweight}, which we abstractly define as follows:

\begin{definition}[Verifiable Distributed Multi-Point Functions (VDMPFs)]
    \textit{Let $L$ be the domain size of the $n$-party multi-point function, $\lambda$ be the security parameter, and $\mathcal{B}$ be the message space. The function scheme consists of three PPT algorithms for all $i \in \{1, 2, ..., n\}$:}
    \begin{itemize}
        \item $\mathsf{Gen}(1^\lambda, \{\alpha_1, \alpha_2, \dots, \alpha_m\} \in \{1, 2, ..., L\}, \{\beta_1, \beta_2, \dots, \beta_m\} \in \mathcal{B}) \rightarrow (k_1, k_2, ..., k_n)$
        \item $\mathsf{Eval}(k_i) \rightarrow (\beta_{i,1}, \beta_{i,2}, ..., \beta_{i,L}, \rho_i)$ 
        \item $\mathsf{Verify}(\rho_1, \rho_2, \dots, \rho_n) \rightarrow 1/0$
    \end{itemize}
    \textit{These algorithms must satisfy the following properties if the VDMPF is secure:}
    \begin{itemize}
        \item \textbf{Correctness.} \textit{A correct VDMPF must ensure that the sum of evaluation outputs equals $\beta_j$ at point $\alpha_j$ for $j \in \{1, 2, ..., m\}$ and $0$ for all other inputs, and that verification succeeds}: 
        $$\Pr\left[
        \begin{array}{cc}
             (k_1, k_2, ..., k_n) \leftarrow \mathsf{Gen}(1^\lambda, \{\alpha_1, \alpha_2, \dots, \alpha_m\}, \{\beta_1, \beta_2, \dots, \beta_m\})  \\
             (\beta_{i,1}, \beta_{i,2}, ..., \beta_{i,L}, \rho_i) \leftarrow \mathsf{Eval}(k_i) \\
             \sum_{i=1}^n \beta_{i,j} = 0,\; j\notin \{\alpha_1, \alpha_2, \dots, \alpha_m\}  \\
             \sum_{i=1}^n \beta_{i,j} = \beta_k,\; j = \alpha_k \text{ for } k \in \{1, 2, \dots, m\} \\
             \mathsf{Verify}(\rho_1, \rho_2, \dots, \rho_n) =1
        \end{array} \right] = 1$$
        \item \textbf{Privacy.} \textit{A VDMPF is private if any strict subset of evaluation keys reveals no information about the inputs, meaning there exists an efficient simulator $\mathsf{Sim}$ that generates outputs computationally indistinguishable from strict subsets of keys output by $\mathsf{Gen}$}.
        \item \textbf{Soundness.} \textit{Soundness guarantees that if verification passes, the evaluated function is well-formed, meaning it has at most $m$ valid points. Formally, for any set of keys, the probability that verification accepts an invalid function (with more than $m$ non-zero positions) is negligible:}
        $$\Pr\left[
        \begin{array}{cc}
            \forall i: (\vec{\beta}_i, \rho_i) \leftarrow \mathsf{Eval}(k_i) \\
            \mathsf{Verify}(\rho_1, \dots, \rho_n) = 1 \\
            \left|\left\{j \in \{1, \dots, L\} : \sum_{i=1}^{n} \beta_{i,j} \neq 0\right\}\right| \leq m
        \end{array} \right] \geq 1 - \mathsf{negl}(\lambda)$$
        \textit{This means that with overwhelming probability, if \(\mathsf{Verify}\) outputs 1, at most $m$ positions have a non-zero sum of the evaluations.}
    \end{itemize}
\end{definition}

After replacing VDPF with VDMPF, we obtain Pepper's final batch messaging protocol. All servers maintain a vector of $L$ verification keys $(g^{\alpha_1}, g^{\alpha_2}, \ldots, g^{\alpha_L})$, where some keys may have the same value, indicating that a public key binds to multiple channels. The directory service publishes the maximum number of sub-messages $t$ that can be sent in a single epoch and the parameter $t$ applies to all broadcasters uniformly. 

For a broadcaster with bandwidth $b$ who needs to write to $t$ of the $b$ channels, with message set $[m] = \{m_1, m_2, \ldots, m_t\}$ and corresponding indices $[j] = \{j_1, j_2, \ldots, j_t\}$, the protocol operates as follows:

\textbf{Step 1: Message Sharing.} Let $y = \alpha_j$ and $[j'] = \{j_1, j_2, \ldots, j_t\}$. Only broadcasters have non-zero messages $m \neq 0$. Each client performs:
\begin{enumerate}
    \item $[m'] \leftarrow \{1^{\text{bit}} \| m_1, 1^{\text{bit}}  \| m_2, \ldots, 1^{\text{bit}}  \| m_t\}$
    \item $(k_1, k_2) \leftarrow \mathrm{VDMPF.Gen}(1^\lambda, [m'], [j'])$
    \item $(\pi_1, \pi_2) \leftarrow \mathrm{SPoSS.Prove}(\alpha_j)$
    \item Encrypt share-pair $(\pi_i, k_i)$ and send to server $i$, for $i \in \{1, 2\}$
\end{enumerate}

\textbf{Step 2: Message Auditing.} After all nodes in the path correctly decrypt the message and server $i$ receives share $(\pi_i, k_i)$ from a client, each server performs:
\begin{enumerate}
    \item $(m_{i}', \rho_i) \leftarrow \mathrm{VDMPF.Eval}(k_i)$
    \item $m_i^\text{bit} \leftarrow \text{Aggregation of first bit of } m'_i[j],\; j \in \{1, \dots, L\}$
    \item $A \leftarrow \sum_{j=1}^L g^{\alpha_j} \cdot m_{i}^{\text{bit}}[j] / t$
    \item $\tau_i \leftarrow \mathrm{SPoSS.Audit}(A, \pi_i)$
    \item Send $(\tau_i, \rho_i)$ to the other server
\end{enumerate}
All servers verify: $\mathrm{SPoSS.Verify}(\tau_1, \tau_2) = 1$ and $\mathrm{VDMPF.Verify}(\rho_1, \rho_2) = 1$. If verification passes, proceed to Step 3. Note that this auditing step processes all $t$ messages in a single verification operation, requiring only one proof token exchange regardless of $t$, in contrast to VDPF-PACLs which would require $t$ separate verifications.

\textbf{Step 3: Message Recovery.} Each server maintains an accumulator $acc_i$ of $L$ entries (channels). The accumulator is defined as $acc_i = \sum_{j \in \mathcal{C}} \mathrm{VDMPF.Eval}(k_j)$, where $\mathcal{C}$ is the set of all clients whose shares passed auditing, and $k_j$ is the VDMPF key from client $j$:
\begin{enumerate}
    \item Compute $(acc_i, \rho_i) \leftarrow \sum_{j \in \mathcal{C}} \mathrm{VDMPF.Eval}(k_j)$
    \item Publicly reveal $acc_i$
\end{enumerate}
Using the publicly revealed shares, anyone can recover the $L$ broadcast messages $m = \sum acc_i$. 

\subsection{Compatibility with Existing Infrastructures and Protocols}
\label{subsec:extension}
In this subsection, we highlight the compatibility of {\pepper} with existing Anonymous Broadcast Protocol (ABP) features and decentralized infrastructures. We discuss how Pepper can seamlessly integrate with blame protocols proposed in prior works, onion routing networks, and federated social networks.

\subject{On-Demand Blame Protocols} Some DC-net protocols, such as Spectrum, include a "blame" mechanism (e.g., the BlameGame subprotocol in Spectrum~\cite{newman2022spectrum}) to identify servers that may have tampered with client requests to falsely accuse a client. A blamed server is subsequently removed from use, while malicious clients are reported and their requests aborted. Although we do not introduce a new blame protocol, Pepper is fully compatible with BlameGame, as Pepper's auditing tokens and message proofs are enhancements of Spectrum's existing features. 

\subject{Co-Deployment with Onion Routing Networks} Pepper's architecture naturally aligns with onion routing infrastructures such as Tor~\cite{dingledine2004tor}. Aggregation servers in Pepper can be co-deployed as part of existing Tor relays, participating in the relay network while simultaneously performing aggregation duties. This dual role enables efficient resource utilization: the same infrastructure that provides anonymous routing can also support anonymous broadcast functionality.

Pepper's directory service can be integrated into Tor's directory service architecture. Specifically, Pepper's directory service can operate as a specialized component within Tor's directory system, publishing information about available aggregation server pairs alongside standard relay information. This integration allows clients to discover Pepper aggregation pairs through the same directory mechanism they use to locate Tor relays, reducing the need for separate infrastructure and improving system discoverability.

\subject{Integration with Federated Social Networks} Pepper's design is compatible with federated social networking architectures, such as Mastodon~\cite{mastodon_project} and other ActivityPub-based systems. In such deployments, Mastodon instances (servers) can serve as Pepper aggregation servers. Each Mastodon instance can establish a pairing relationship with another instance, forming aggregation pairs that process anonymous broadcast messages while continuing to handle standard federated social networking traffic.

Ordinary users in federated social networks can naturally contribute cover traffic for Pepper. When users interact with the social network (e.g., posting, liking, or following), their network activity can be leveraged to generate cover traffic for Pepper's anonymous broadcast channels. This dual-purpose traffic helps maintain anonymity sets without requiring users to explicitly participate in cover traffic generation. Additionally, some Mastodon instances can be designated to generate cover traffic, further enhancing anonymity guarantees while utilizing existing server infrastructure.

This compatibility enables practical deployment scenarios where anonymous broadcast functionality is seamlessly integrated into existing decentralized social platforms. Users benefit from enhanced privacy-preserving communication capabilities without requiring separate infrastructure or significant modifications to existing protocols.

Next, we implement and evaluate {\pepper}, respectively explore the communication efficiency (Section.\ref{subsec:computation}), Computation Overhead (Section.\ref{subsec:overhead}), thoughput (\ref{subsec:throughput}) and end-to-end transfer latency (Section.\ref{subsec:latency}) of {\pepper}. 

\section{Security and Performance Analysis}
\label{sec:analysis}

This section analyzes the theoretical efficiency and security of Pepper with respect to the registration and messaging subprotocols in Section~\ref{subsec:registration} and Section~\ref{subsec:messaging}. We use $\lambda$ for the system-configured security parameter, $N$ for the number of clients participating in an epoch, and $L$ for the number of channels participating in the current protocol stage, namely registration channels during registration and message channels during messaging. Let $|m|$ denote the bit length of a registration record or a single message payload, and let $t$ denote the maximum number of message-channel writes encoded in one batch messaging request.

\subsection{Efficiency Analysis}
\label{subsec:efficiency-analysis}

\subject{Communication Efficiency.} In registration, a client sends one VDPF key share to each aggregation server. The per-server client-to-server request size is $O(\lambda \log L + |m|)$, including the VDPF key material for a registration record. During auditing, each server exchanges an $O(\lambda)$ VDPF verification token with the other server. After accepting valid registration requests, each server reveals or exchanges an accumulator over $L$ registration channels, which costs $O(L |m|)$ per registration epoch.

In messaging, a client sends one VDMPF key share and one SPoSS proof share to each server. Following the Big-State DMPF construction used by Pepper~\cite{boyle2025improved}, the per-server VDMPF key size is $O((\lambda t + t^2)\log L + t|m|)$, while the SPoSS proof share is $O(\lambda)$~\cite{servan2023private}. The audit requires only one VDMPF verification token and one SPoSS token exchange per batch, for $O(\lambda)$ inter-server communication independent of $t$. Finally, each server reveals an accumulator over the $L$ message channels, costing $O(L|m|)$ per messaging epoch. Thus, for $N$ clients, the audit-token traffic scales as $O(N\lambda)$, while the final aggregation traffic is independent of $N$ once all valid requests have been accumulated.

\subject{Computational Efficiency.} In registration, each client generates a VDPF key in $O(\lambda \log L + |m|)$ time. Each server expands and verifies the received VDPF share, producing both an $L$-entry share vector and a verification token. This costs $O(\lambda \log L + L|m|)$ work per client request when accounting for vector expansion and accumulation. Across an epoch, server-side registration work is $O(N(\lambda \log L + L|m|))$.

In messaging, each client generates a SPoSS proof share in $O(\lambda)$ time and a VDMPF key in $O(t^2\lambda \log L + t|m|)$ time~\cite{boyle2025improved}. Each server evaluates the VDMPF key and computes the SPoSS accumulator over the verification-key vector, yielding $O(t^2\lambda \log L + t|m| )$ work per client request. The important distinction from applying VDPF-PACLs independently $t$ times~\cite{servan2023private} is that Pepper performs one batch audit and exchanges one pair of audit tokens for the whole $t$-message request. The per-epoch server work is therefore $O(N(t^2\lambda \log L + t|m|))$.

\subsection{Security Analysis}
\label{subsec:security-analysis}

We analyze Pepper in the standard DC-net setting where at least one aggregation server in the selected cohort is honest. The adversary may observe network traffic, corrupt any strict subset of servers, and corrupt arbitrary clients. Client-to-server traffic is encrypted and padded to the public request size for the corresponding protocol phase. We rely on the privacy and soundness of VDPF/VDMPF constructions~\cite{boyle2016function,de2022lightweight,boyle2025improved}, the soundness and zero-knowledge properties of SPoSS~\cite{servan2023private}, collision-resistant hashing, and the assumptions listed in Section~\ref{sec:overview}.

\subject{Correctness.} For honest clients and honest execution by the servers, registration correctness follows from VDPF correctness and the final hash check on recovered registration records. A non-colliding registration request reconstructs exactly one record $g^\alpha \| l_2 \| b \| \mathcal{H}(g^\alpha \| l_2 \| b)$, and servers bind the requested message-channel range to $g^\alpha$. If multiple clients collide on the same registration slot, the protocol discards the recovered value and the client retries in a later epoch. Messaging correctness follows from VDMPF correctness and SPoSS completeness. An honest broadcaster encodes at most $t$ non-zero payloads at registered channels, the batch audit accepts, and the public sum of server accumulators reconstructs the intended messages. Honest cover-traffic clients either write zero or target phantom channels whose outputs are ignored by the bulletin, so they do not alter real message channels.

\subject{Anonymity.} Consider two honest clients in the same epoch, one acting as a broadcaster and the other contributing cover traffic. For any adversary corrupting only a strict subset of servers, the VDPF or VDMPF key shares seen by corrupted servers can be simulated from the public parameters by the privacy of the underlying FSS primitive. SPoSS proof shares are zero-knowledge, and the exchanged audit tokens reveal only whether the request passed verification. Since encrypted request payloads have the same public size within a protocol phase, the adversary's view is computationally indistinguishable between the broadcaster and cover-traffic client, except for leakage inherent in the final public bulletin and participation patterns. If all aggregation servers in a cohort collude, Pepper does not claim cryptographic sender anonymity for that cohort.

\subject{Registration Soundness.} A malicious client should not be able to register malformed data or bind multiple registration slots through one accepted request. If a registration request passes auditing, VDPF soundness implies that the reconstructed request has at most one non-zero registration-channel entry except with negligible probability in $\lambda$. The collision-resistant hash binds the recovered public key, starting message channel, and bandwidth value. Servers reject records with invalid hashes or overlapping message-channel ranges. Therefore, except with negligible probability, every accepted registration is either a single well-formed binding or an explicitly detected collision that is discarded.

\subject{Message Access-Control Soundness.} A malicious client should not be able to write a non-zero message to a registered channel without knowing the corresponding secret key, nor write more than the configured batch size. If a messaging request passes VDMPF verification, VDMPF soundness implies that the request encodes at most $t$ non-zero channel positions except with negligible probability. If the SPoSS audit accepts, SPoSS soundness implies that the client knows a discrete logarithm consistent with the verification-key accumulator used in the audit. The checksum and accumulator construction bind the non-zero payload positions to the registered verification keys. Thus, for batch requests targeting channels bound to the same broadcaster key, an accepted non-zero write is authorized by knowledge of the corresponding private key, unless the adversary breaks VDMPF soundness, SPoSS soundness, collision resistance, or the underlying discrete-logarithm assumption. Writes to phantom channels are permitted only as cover traffic and are ignored during bulletin publication.

\subject{Malicious Servers and Blame.} The arguments above address privacy and request validity when at least one server follows the protocol and accepted transcripts are processed as specified. Pepper does not introduce a new complete blame proof. Instead, as discussed in Section~\ref{subsec:extension}, Pepper is compatible with Spectrum's BlameGame~\cite{newman2022spectrum}: if a server tampers with client requests or falsely attributes an audit failure, the deployment can use the same on-demand blame approach to attribute misbehavior. This compatibility does not eliminate denial-of-service by malicious servers, which remains outside Pepper's core security goals.

\section{Implementation and Evaluation}
\label{sec:eval}








In this section we present the implementation and evaluation of Pepper. 
We conduct experiments in a standard 2-server DC-net setting and compare Pepper against three representative systems: Express~\cite{eskandarian2021express}, Spectrum~\cite{newman2022spectrum}, and PACLs~\cite{servan2023private}. In our evaluation, we focus on analyzing and profiling the following aspects of Pepper: (i) Communication efficiency; (ii) Computation overhead; (iii) End-to-end broadcasting throughput. 





\textbf{Implementation.} We implemented Pepper with approximately 8,000 lines of Go (Go 1.24) code for Pepper client and server, 2,900 lines of C/C++ code for the VDMPF library construction. Our (V)DMPF implementation follows the Big-State DMPF~\cite{boyle2025improved}, the state-of-the-art construction for batch message size $t < 70$. Regarding the three baseline protocols, we utilized the reference implementations~\cite{ExpressRepo,SpectrumImpl, PACLRepo} as released by the respective authors. Across these implementations, we use the P-256 elliptic curve group for Spectrum group $\mathbb{G}$ construction. For VDPF-PACLs and Pepper, we initiate $\mathbb{G}$ as field $\mathbb{F}_p^*$ with 3072-bit prime $p$ specified in RFC3526~\cite{rfc3526}. Across these implementations, traffic is encrypted with TLS 1.3.


\textbf{Environment}. Client–server latency is measured using a 24-core 2.2 GHz Intel Xeon server (64 GB RAM, acting as client) and a 24-core 2.4 GHz Intel Xeon server (128 GB RAM, acting as server), both running Ubuntu 22.04 and located in the same data center to reduce network fluctuation. All other experiments are conducted on a workstation with an Intel i5-13600KF CPU and 32 GB RAM running Ubuntu 24.04, utilizing AES-NI for efficient PRG evaluation and standard D(M)PF optimizations; turbo-boost is disabled to minimize performance variance. Servers are pinned to distinct CPU cores while a single core emulates all clients.


\textbf{Result Calculation}. We repeat each experiment between 3 and 100 times, depending on the experiment. For throughput and latency benchmarks, we perform one additional warm-up trial that is not included in the reported results. To mitigate the influence of outliers, we compute a 5\% trimmed mean by discarding the top and bottom 5\% of the data.


\subsection{Communication Cost}
\label{subsec:overhead}

\begin{table}
    \centering
    \caption{Communication cost measured by payload size. All experiments are configured with the security parameter $\lambda=128$ and the channel size $L=1024$. The payload types include Proof (pre-audit token, client$\to$server), Key (FSS key size, client$\to$server), and Audit (verification token, server$\leftrightarrow$server). For messaging, $t$ is the batch messaging size. 
    }
    \label{tab:communication_cost_stacked}
    \begin{tabular}{lccc}
        \toprule
        & \multicolumn{2}{c}{\textbf{client$\to$server}} & \textbf{Server$\leftrightarrow$Server} \\
        \cmidrule(lr){2-3} \cmidrule(lr){4-4}
        \textbf{Registration} & Proof & Key & Audit \\
        \midrule
        Pepper & - & $|m| + 262$ bytes & 16 bytes \\
        Spectrum & 64 bytes & $|m| + 198$ bytes & 64 bytes \\
        Express & 2304 bytes & $|m| + 198$ bytes & 2304 bytes \\
        \midrule
        \multicolumn{4}{c}{} \\[-1ex]
        \textbf{Messaging} & Proof & Key & Audit \\
        \midrule
        Pepper & 1952 bytes & $(|m| + 304)t + 64$ bytes & 816 bytes \\
        Spectrum & $64t$ bytes & $(|m| + 198)t$ bytes & $64t$ bytes \\
        PACLs & $1952t$ bytes & $(|m| + 262)t$ bytes & $816t$ bytes \\
        \bottomrule
    \end{tabular}
\end{table}

The evaluation distinguishes between two protocol phases with different security requirements. The registration protocol only needs to verify that messages are well-formed, while the messaging protocol additionally requires verifying that the client possesses the corresponding private key (or is sending cover traffic). Express, Spectrum, and PACLs all implement algorithms for the well-formedness verification. However, Express cannot verify a broadcaster's authorization to a specific channel. Since Pepper's registration protocol follows the same approach as PACLs, we compare Express, Spectrum, and Pepper for registration. For messaging, we compare Spectrum, PACLs with Pepper, as these systems all implement access control mechanisms.

Table~\ref{tab:communication_cost_stacked} reports communication payload in bytes grouped by protocol phases (registration and messaging), communication parties (Client-Server and Server-Server), and payload purposes, e.g., ownership proof of the channel private key (Proof) from the client, the distributed point function key (Key) from the client, and the auditing tokens exchanged among servers.

All numeric values are calculated under the security parameter of $\lambda=128$, channel number $L = 1024$ and $|m|$-bit messages.  For messaging, both of the Client-Server and Server-Server payloads in Spectrum and PACLs involves a multiplicative factor of $t$, because $t$ epochs are needed in these systems to broadcast a $t$-message batch. Key sizes are similar across systems; Pepper's messaging DPF keys are modestly larger because Big-State DMPF trades slightly larger keys for faster reconstruction. Spectrum has the smallest proof and audit tokens when $t$ is small, whereas Pepper attains competitive totals while enabling single-shot auditing for a batch of $t$ messages.

\subsection{Auditing Time}
\label{subsec:computation}
This subsection measures the time servers spend executing the verification procedure (auditing) for each client request. We report per-request auditing time, which is the CPU time required for servers to verify DPF proof tokens, check channel write authorization, and validate message shares before aggregation. This metric directly determines auditing throughput (audits/sec) and bounds end-to-end throughput when CPU is the bottleneck.

\begin{figure}
    \centering
    \includegraphics[width=0.48\linewidth]{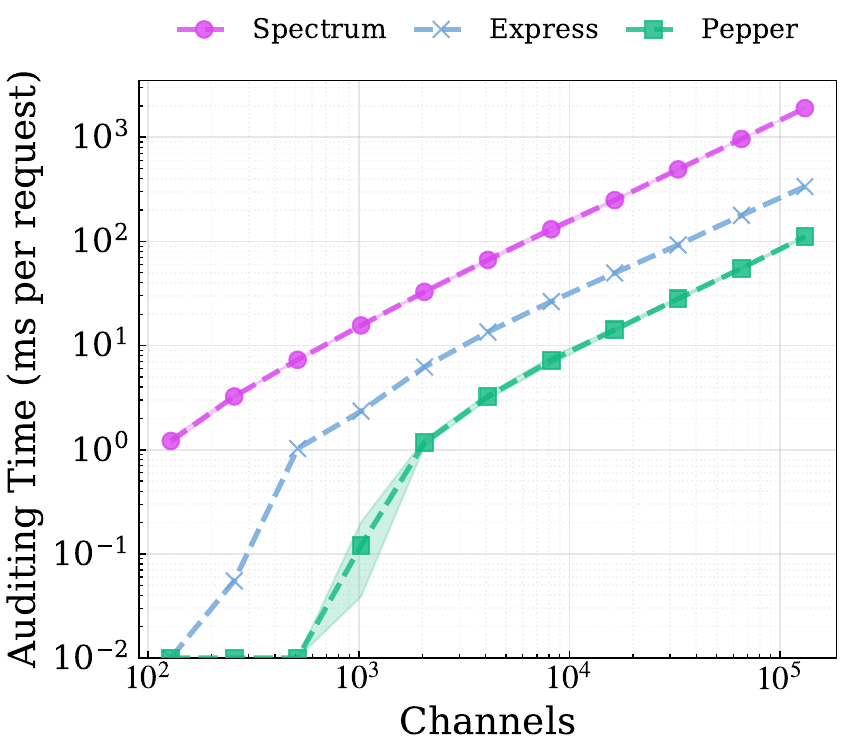}
    \hfill
    \includegraphics[width=0.48\linewidth]{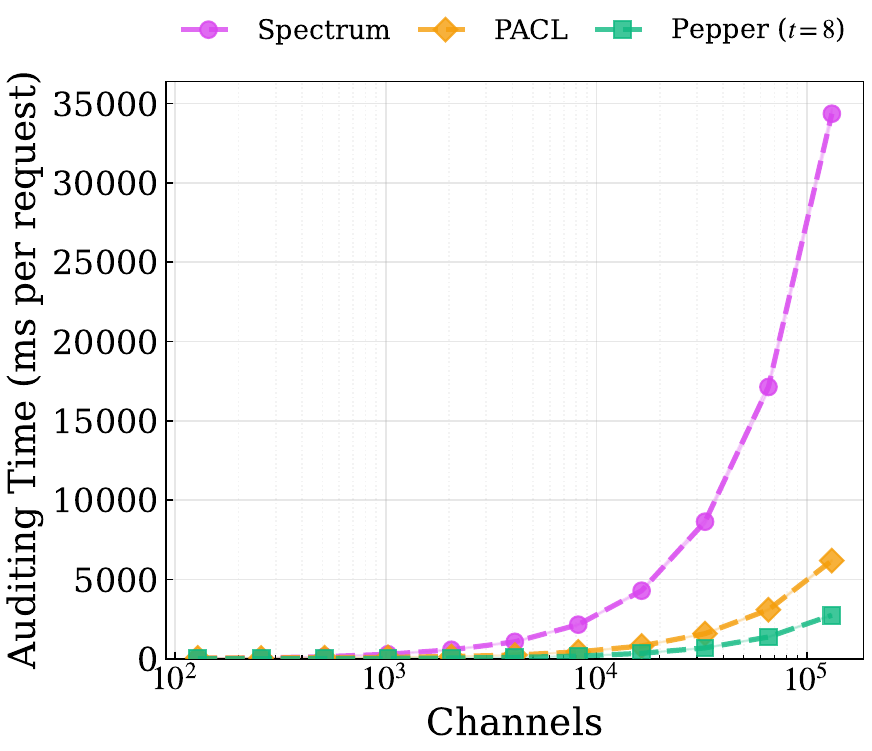}
    \caption{Per-request auditing time (lower is better). Left: registration. Right: messaging, with total message size 5KB and message batch size $t=8$. Pepper benefits from (1) lightweight audit-share generation and proof computation in registration, and (2) batch-style auditing for batch messaging.} 
    \label{fig:computation_cost}
\end{figure}

\subject{Registration.} As the number of channels increases, auditing time grows for all systems but at different rates (Figure\,\ref{fig:computation_cost}, left). Spectrum is dominated by group operations (exponentiations and multiplications). By contrast, Pepper performs a single VDPF expansion for hash verification and achieves the lowest per-request auditing time. Across the evaluated channel range (1K to 16K), Express is between $3.0\times$ and $19.5\times$ slower than Pepper, while Spectrum is between $17.2\times$ and $130\times$ slower. These per-auditing-request times determine the server-side capacity that bounds throughput when CPU is the bottleneck.

\subject{Messaging.} For messaging (Figure\,\ref{fig:computation_cost}, right), Spectrum is fast at small channel counts (e.g., $<\!500$) due to its lightweight auditing token design, but its auditing time grows rapidly by the number of channels. PACLs reduce exponentiations to multiplications (and multiplications to additions), improving scalability over Spectrum. Building on VDPF-PACLs auditing, Pepper verifies a batch of $t$ messages (total size keeps same to VDPF-PACLs and Spectrum) in one shot, without increasing the audit token size, yielding better scaling than Spectrum and competitive auditing time versus PACLs. 
For 5\,KB per-client messages with $t{=}8$, across the evaluated channel range (1K to 16K), PACLs are between $2.3\times$ and $3.6\times$ slower than Pepper, while Spectrum is between $10.8\times$ and $12.4\times$ slower. These measurements are per auditing request and translate directly to throughput when computation is the bottleneck.

\subsection{Throughput}
\label{subsec:throughput}
The real-world throughput of an anonymous broadcast system is influenced by various factors, including the number of broadcasters and cover-traffic clients per epoch, the maximum number of messages per client when batch messaging is supported, the computing capacity of aggregation servers, and network bandwidth and congestion conditions. An exhaustive evaluation of all these factors is impractical. Instead, we focus on stress-testing the processing capability of the servers, as this is typically the bottleneck for throughput in real-world deployments.

To profile the server's processing throughput, we report two complementary metrics. The first is \emph{client-request throughput} (requests/sec), which represents the maximum number of client requests that a typical server deployment can process. Each client request, whether it is cover traffic or a true broadcast request, is processed by the server cohort through a sequence of steps, including DPF key evaluation, proof verification, and local access token computation. In our experiments, we measure how this throughput metric varies with the number of channels, the extent of message batching, and the message size.

The second metric is \emph{auditing-request throughput} (audits/sec), which captures the maximum number of audits the servers can perform per second. An auditing request involves executing the servers' verification steps for a given client request, with the goal of detecting potential client disruptions. 
As auditing is the most computation-intensive step in client request processing, making this metric particularly relevant.

\begin{figure}
    \centering
    \includegraphics[width=\textwidth]{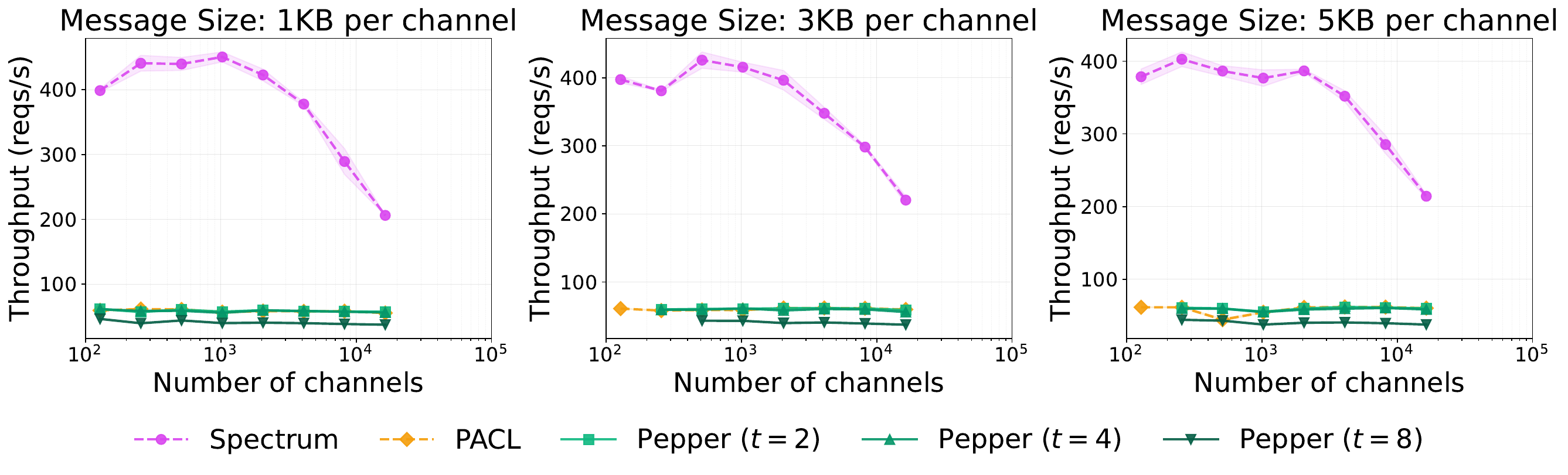}
\caption{Client-request throughput measured by the number of client requests processed per second by aggregation servers. The three panels correspond to different per-channel message sizes: 1\,KB, 3\,KB, and 5\,KB, respectively. For Pepper, the throughput is further analyzed as a function of the batch messaging size $t\in\{2,4,8\}$.}
    \label{fig:throughput_client_triptych}
\end{figure}

Figure~\ref{fig:throughput_client_triptych} presents the client-request throughput as a function of message sizes (1\,KB, 3\,KB, and 5\,KB) and protocols. Notably, we report throughput results only for the messaging subprotocol, as registration is typically a one-time process involving a single epoch, whereas messaging, as the core of Pepper, spans numerous epochs and is more performance-critical. Limiting the evaluation to the messaging subprotocol also facilitates direct comparison with the three baseline systems, as all of them specify only the messaging subprotocol. As shown in Figure~\ref{fig:throughput_client_triptych}, Spectrum achieves higher client-request throughput ($3.6$--$7.8\times$ that of Pepper) because it is optimized for scenarios where each request contains a single message. In contrast, Pepper is designed for batch messaging, where multiple messages are processed per request, resulting in higher computation costs per client request and, consequently, lower client-request throughput.

\begin{figure}
    \centering
    \includegraphics[width=\textwidth]{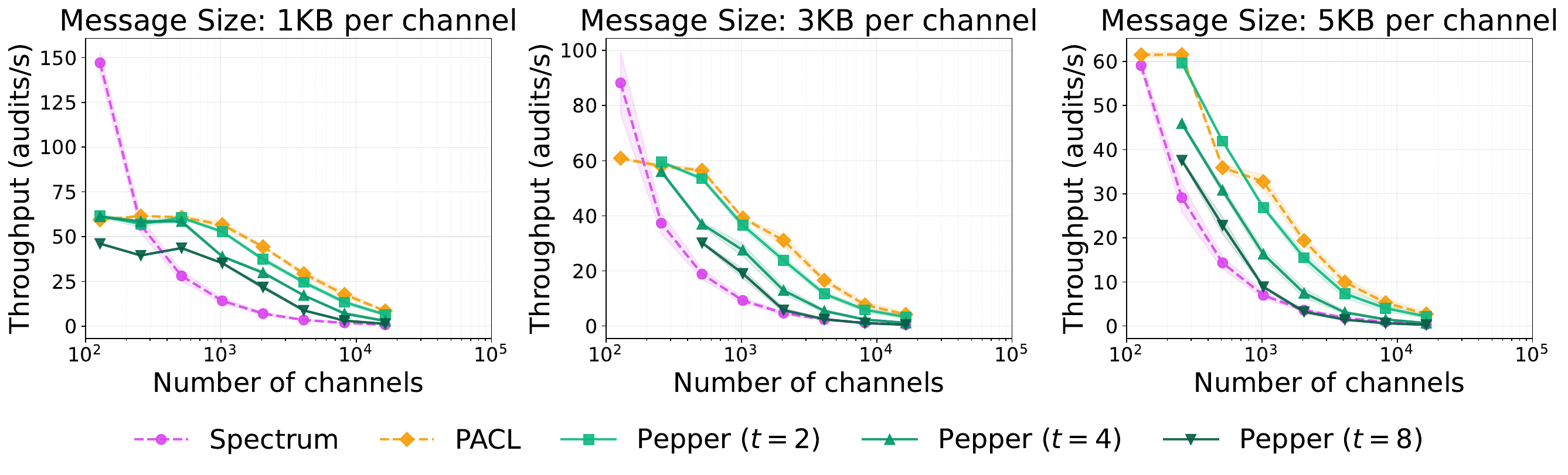}
\caption{Auditing throughput measured by the number of audits processed per second by aggregation servers. The three panels correspond to different per-channel message sizes: 1\,KB, 3\,KB, and 5\,KB, respectively. For Pepper, the throughput is further profiled as a function of the batch messaging size $t\in\{2,4,8\}$, highlighting its scalability with batch size.}
    \label{fig:throughput_audit_triptych}
\end{figure}

However, when considering auditing-request throughput, as shown in Figure~\ref{fig:throughput_audit_triptych}, Pepper with batching sizes $t\in\{2,4,8\}$ achieves performance comparable to PACLs and often surpasses Spectrum, particularly when the number of channels is practically large (e.g., over $10^3$). This demonstrates that, under the same computing constraints, Pepper can achieve comparable or higher auditing throughput while delivering $t\times$ messages per audit. 

Particularly, our further experiments show that at $t{=}8$, Pepper achieved between $1.2\times$ and $5.0\times$ the effective messaging rate of PACLs and between $13.6\times$ and $20\times$ that of Spectrum across the evaluated channel range, highlighting the advantages of batch messaging.

\subsection{Latency}
\label{subsec:latency}
This subsection measures latency of Pepper along with a direct comparison with baselines. We run VMs on Amazon EC2 \texttt{c5.4xlarge} instances with 8 cores and 32 GB RAM to simulate a real-world WAN deployment. Each instance runs on Ubuntu 20.04. We run clients in Virginia while servers are located in Oregon and Ohio. The TLS connection round trip times (RTTs) were 26.01 ms between Ohio and Oregon, 105.32 ms between Ohio and Virginia and 24.11 ms between Virginia and Oregon. 


\subject{End-to-End File Transmission Latency.} We first evaluate the end-to-end latency of Pepper, defined as the time required to broadcast a file of practical sizes. In this evaluation, we fix Pepper's batch messaging size to $t{=}4$ and compare it against Spectrum and PACLs under identical settings, including 2,048 channels and 100 clients. We select files of four practical sizes: 100\,KB, 1\,MB, 10\,MB, and 100\,MB.

Table~\ref{tab:file_transfer} presents the average end-to-end latency for different file sizes and protocols. Pepper consistently achieves the lowest latency across all measured file sizes, with PACLs being $1.32$--$1.63\times$ slower and Spectrum being $3.77$--$26.90\times$ slower than Pepper. Because Spectrum's 100\,MB experiment would take prohibitively long to complete, we estimate it from single-message processing logs, each Spectrum 100\,MB message takes $2.71$\,hours on average to process, and the reported value extrapolates this cost to 100 clients before adding the merge latency. It is important to note that real-world latency is influenced by numerous factors; thus, the latency values reported in Table~\ref{tab:file_transfer} are intended for comparative analysis between Pepper and the baseline protocols.

\begin{table}
\centering
\caption{End-to-end file transmission latency (lower is better). Pepper uses a batch messaging size of $t{=}4$. Each cell reports the average latency $\pm$ standard deviation. Latency is measured from client start to server stop, plus merge duration. Spectrum's 100\,MB result is estimated from single-message processing logs.}
\label{tab:file_transfer}
\begin{tabular}{lccc}
\toprule
\textbf{File Size} & \textbf{Pepper} & \textbf{PACLs} & \textbf{Spectrum} \\
\midrule
100\,KB & $33.18 \pm 0.29\,\text{s}$ & $53.23 \pm 0.52\,\text{s}$ & $2.08 \pm 0.01\,\text{min}$ \\
1\,MB & $11.42 \pm 0.05\,\text{min}$ & $15.12 \pm 0.03\,\text{min}$ & $49.30 \pm 0.07\,\text{min}$ \\
10\,MB & $1.79 \pm 0.02\,\text{h}$ & $2.53 \pm 0.06\,\text{h}$ & $8.40 \pm 0.04\,\text{h}$ \\
100\,MB & $10.20 \pm 0.02\,\text{h}$ & $16.68 \pm 0.22\,\text{h}$ & $274.42 \pm 0.08\,\text{h}$ \\
\bottomrule
\end{tabular}
\end{table}

\subject{Latency Composition.} To better understand the factors contributing to overall latency, we analyzed the latency composition by leveraging fine-grained performance logging. This allowed us to break down the total latency into distinct components, including client-server communication latency, server-server communication latency, and computation latencies for both the client and server.

\begin{figure}
    \centering
    \begin{minipage}[t]{0.48\textwidth}
        \centering
        \includegraphics[width=\linewidth]{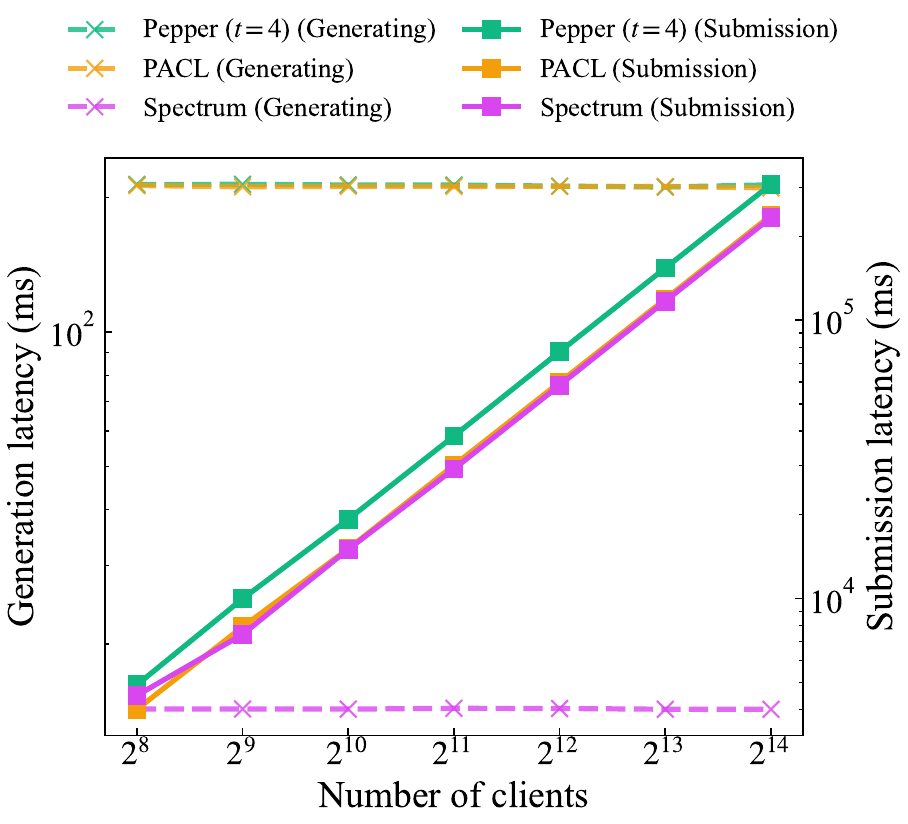}
        \caption{Client-server latency (lower is better) of the messaging process. The channel size is 2,048, and the per-client message size is 10\,KB. The left y-axis reports generation latency, and the right y-axis reports submission latency.}
        \label{fig:cs_latency}
    \end{minipage}
    \hfill
    \begin{minipage}[t]{0.48\textwidth}
        \centering
        \includegraphics[width=\linewidth]{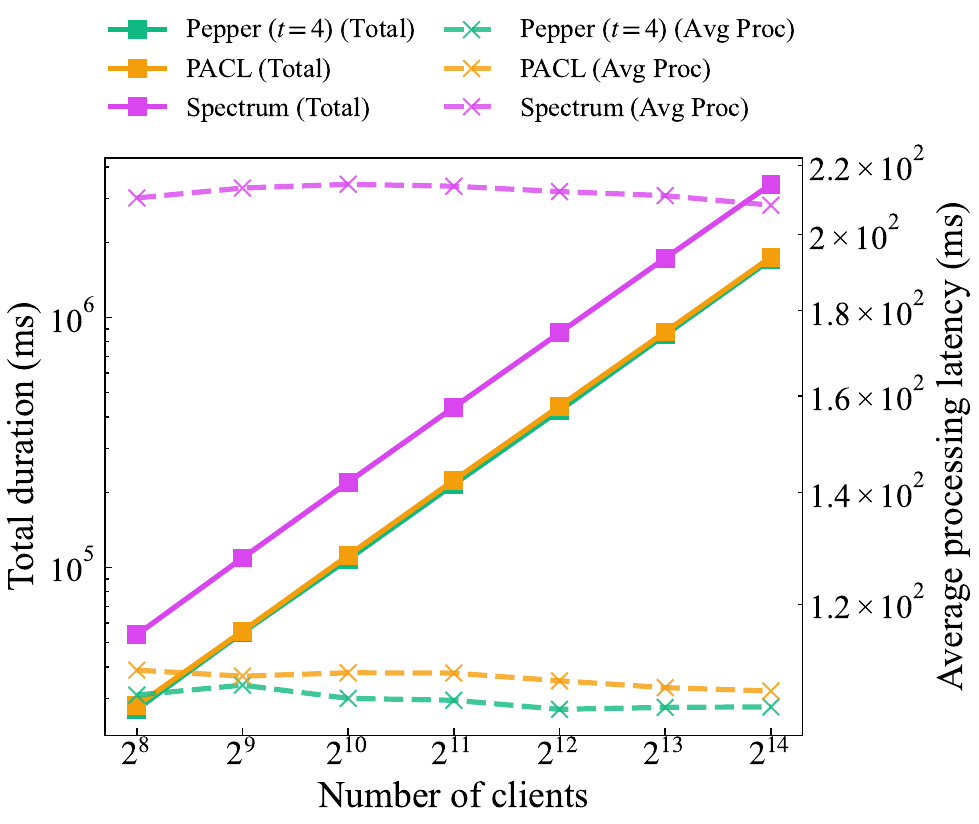}
        \caption{Server-server latency (lower is better) of the messaging process. The channel size is 2,048, and the per-client message size is 10\,KB. The left y-axis reports total duration, and the right y-axis reports average processing latency.}
        \label{fig:ss_latency}
    \end{minipage}
\end{figure}

\subsubject{Client-Server Latency.} Figure~\ref{fig:cs_latency} reports the client-server latency for messaging. Two metrics are evaluated: \emph{generation latency}, which is the total time for a client to generate and prepare a request (including proof computation and key generation), and \emph{submission latency}, which is the time from when the first message is sent until the last message is received by the server. For the evaluated message sizes, generation latency and submission latency are nearly identical because network transmission time is negligible compared to cryptographic computation time. Pepper, with $t{=}4$, is $14.8$--$15.0\times$ slower than Spectrum and $1.00$--$1.02\times$ slower than PACLs due to the higher cost of proof generation for batch messaging. In terms of submission latency, Pepper is $1.09$--$1.34\times$ slower than Spectrum and $1.22$--$1.29\times$ slower than PACLs. However, this overhead is small in absolute terms and can be incurred offline before submission.

\subsubject{Server-Server Latency.} We measure two metrics for server-server latency: \emph{total duration}, which is the time from processing the first message to completing the merge, and \emph{average processing latency}, which is the average time per message. The gap between these metrics does not scale linearly with the number of clients due to parallel processing. As shown in Figure~\ref{fig:ss_latency}, Spectrum is $1.99$--$2.05\times$ slower than Pepper with $t{=}4$, while PACLs are $1.01$--$1.04\times$ slower. These results highlight Pepper's efficiency in server-server communication, particularly for batch messaging.

\section{Discussion}
\label{sec:discuss}

This section discusses potential future work directions and extensions to the Pepper system. While our current design provides strong anonymity guarantees and efficient batch messaging, several areas warrant further investigation to enhance the system's capabilities and applicability.

\textbf{Multi-Public-Key Batch Messaging.} In our current design, a broadcaster can register multiple channels under a single public key and send a batch of messages to these channels in one request. A promising extension would be to support scenarios where a single broadcaster holds multiple public keys (each corresponding to different channels) and wishes to send a batch of messages in a single request, with each message targeting a channel associated with a different public key. This would enable more flexible communication patterns while maintaining sender anonymity. 


\textbf{Adaptive Channel Allocation.} Currently, Pepper uses a fixed channel allocation strategy where broadcasters register channels during the registration phase. Future work could explore dynamic channel allocation mechanisms that adapt to network conditions, user demand, and security requirements. This might involve developing reputation systems for server pairs and implementing load balancing strategies that maintain anonymity while optimizing performance.

\textbf{Enhanced Cover Traffic Strategies.} While our current phantom channel approach provides basic cover traffic support, more sophisticated strategies could improve anonymity guarantees. Future work could investigate adaptive cover traffic generation that responds to network conditions, user behavior patterns, and potential attacks. This might include developing machine learning-based approaches to generate realistic cover traffic patterns.
\section{Conclusion}
\label{sec:conclusion}

Pepper is a practical anonymous broadcast system for Tor-like settings with multiple server pairs. It combines FSS-based anonymous writes with verifiable auditing and single-shot batch messaging that verifies $t$ messages per request without increasing the audit token size, yielding compact client bandwidth and low auditing CPU cost. Our implementation shows millisecond-level registration audits up to 16K channels and competitive or superior delivered-message throughput at larger channel counts by amortizing verification across $t$ messages. Pepper preserves DC-net anonymity, is compatible with existing blame protocols, and provides a concrete path to higher throughput and lower verification cost in realistic deployments.


\bibliographystyle{IEEEtranS}
\bibliography{ref}

\appendix
\section{Proof of VDMPF-PACLs Collision Probability}
\label{appendix:dmpf_proof}
Let $p$ be a large prime modulus such that the Discrete Logarithm Problem (DLP) in the multiplicative group $\mathbb{Z}_p^*$ is computationally hard, providing a $\lambda$-bit security level. Let $g$ be a generator of a large subgroup of $\mathbb{Z}_p^*$.

Let $\mathcal{P} \subset \mathbb{Z}_p$ be the set of all registered public keys, where each $P_k \in \mathcal{P}$ is of the form $P_k = g^{\alpha_k} \pmod p$ for some private key $\alpha_k$. Consider a malicious client with public key $P_x = g^{\alpha_x} \pmod p$ who defines a target value $V = (t \cdot P_x) \pmod p$ where $t \in \mathbb{Z}^+$.

The client selects a $t$-multiset $\mathcal{S} = \{P_1, P_2, \dots, P_t\}$ from $\mathcal{P}$ under the constraint $\mathcal{S} \neq \{P_x, P_x, \dots, P_x\}$ and computes:
$$S = \left(\sum_{i=1}^{t} P_i\right) \pmod p$$

We seek to determine $\Pr(S = V)$.

Assume that the public key values $P_k = g^{\alpha_k} \pmod p$ are computationally indistinguishable from values drawn uniformly at random from $\mathbb{Z}_p$.

Under the above assumption, the sum $S$ is approximately uniformly distributed over $\mathbb{Z}_p$. Since $\mathbb{Z}_p$ contains $p$ distinct elements, we have:
$$\Pr(S = v) \approx \frac{1}{p}, \quad \forall v \in \mathbb{Z}_p$$

Therefore:
$$\Pr(S = V) = \frac{1}{p}$$

Since $p \geq 2^{\lambda}$ for security parameter $\lambda$, the probability $\frac{1}{p}$ is cryptographically negligible. Thus, it is computationally infeasible for a malicious client to find a non-trivial combination $\mathcal{S}$ of $t$ public keys whose sum equals the target value $V$.

\section{Schnorr Proof over Secret Shares (SPoSS)}
\label{appendix:sposs}

This section provides an algorithmic description of the Schnorr Proof over Secret Shares (SPoSS) primitive used in the VDPF-PACLs construction (Section~\ref{subsub:pacl}). Recall that SPoSS allows a prover to convince two verifiers that it knows the discrete logarithm $x$ of a group element $w = g^x$ while $w$ itself is additively secret-shared between the verifiers.

We work over a cyclic group $\mathbb{G} = \langle g \rangle \subseteq \mathbb{Z}_p^*$ of prime order $p$, and all secret sharing is over the field $\mathbb{F}_p$. Let $w_1,w_2 \in \mathbb{G}$ denote additive shares of $w$ such that $w = w_1 + w_2$ (as in the main text), and let $H$ be a random oracle $H : \{0,1\}^* \rightarrow \mathbb{Z}_p$.

\begin{algorithm}[htp]
\caption{Schnorr Proof over Secret Shares (SPoSS)}
\label{alg:sposs}
\DontPrintSemicolon
\KwIn{Secret exponent $x \in \mathbb{Z}_p$; generator $g \in \mathbb{G}$}
\KwOut{Proof shares $(\pi_1,\pi_2)$; audit tokens $(\tau_1,\tau_2)$; verification bit $b$}
\BlankLine
\subject{SPoSS.Prove$(x)$} \tcp*{prover}
Set $w \gets g^x \in \mathbb{G}$.\;
Sample $r \xleftarrow{\$} \mathbb{Z}_p$ and set $t \gets g^r$.\;
Set $c \gets H(g,w,t) \in \mathbb{Z}_p$.\;
Set $s \gets r + c x \bmod p$.\;
Choose additive shares $(w_1,w_2)$, $(t_1,t_2)$, $(s_1,s_2)$ over $\mathbb{F}_p$ such that
$w = w_1 + w_2$, $t = t_1 + t_2$, $s = s_1 + s_2$.\;
For each $i \in \{1,2\}$, define $\pi_i \gets (w_i,t_i,s_i,c)$ and send $\pi_i$ to verifier $i$.\;
\BlankLine
\subject{SPoSS.Audit$(w_i,\pi_i)$} \tcp*{verifier $i$}
Input local share $w_i$ and proof share $\pi_i = (w_i,t_i,s_i,c)$.\;
Check that $w_i,t_i \in \mathbb{G}$; if a check fails, output a rejecting token.\;
Set $\tau_i \gets (w_i,t_i,s_i,c)$ and send $\tau_i$ to the other verifier.\;
\BlankLine
\subject{SPoSS.Verify$(\tau_1,\tau_2)$} \tcp*{joint verification}
Parse $\tau_i$ as $(w_i,t_i,s_i,c)$ for $i \in \{1,2\}$.\;
Reconstruct $w \gets w_1 + w_2$, $t \gets t_1 + t_2$, and $s \gets s_1 + s_2 \bmod p$.\;
Set $c' \gets H(g,w,t)$.\;
If $c' = c$ and $g^{s} = t \cdot w^{c}$ in $\mathbb{G}$, set $b \gets 1$; otherwise set $b \gets 0$.\;
Return $b$.\;
\end{algorithm}

By construction, if both verifiers hold additive shares $w_1,w_2$ such that $w_1 + w_2 = g^x$ for some $x \in \mathbb{Z}_p$, then an honest prover running $\mathsf{SPoSS.Prove}$ produces proof shares that cause $\mathsf{SPoSS.Verify}$ to output $1$ except with negligible probability. This aligns with the interface used in the main text, where $\mathsf{SPoSS.Verify}(\tau_1,\tau_2)$ returns $1$ if and only if $w_1 + w_2 = g^x$ over $\mathbb{F}_p$.

The SPoSS primitive is required to satisfy the following properties in the random-oracle model (with $H$ as a random oracle).

\paragraph*{Completeness.}
For all $x \in \mathbb{Z}_p$ and $w = g^x \in \mathbb{G}$, let $(\pi_1,\pi_2) \leftarrow \mathsf{SPoSS.Prove}(x)$ and, for $i \in \{1,2\}$, let $\tau_i \leftarrow \mathsf{SPoSS.Audit}(w_i,\pi_i)$ where $w_1,w_2$ are the additive shares of $w$ produced inside $\mathsf{SPoSS.Prove}$. Then
\[
\Pr\left[
\mathsf{SPoSS.Verify}(\tau_1,\tau_2) = 1
\right] = 1,
\]
where the probability is over the randomness of $\mathsf{SPoSS.Prove}$.

\paragraph*{Argument-of-knowledge.}
There exists a probabilistic polynomial-time extractor $\mathcal{E}$ such that the following holds. Let $P^*$ be any (possibly malicious) prover that on input $g$ and $w \in \mathbb{G}$ produces proof shares $(\pi_1^*,\pi_2^*)$ so that, after running
\[
\tau_i^* \leftarrow \mathsf{SPoSS.Audit}(w_i,\pi_i^*) \quad (i \in \{1,2\}),
\]
we have
\[
\Pr\left[
\mathsf{SPoSS.Verify}(\tau_1^*,\tau_2^*) = 1
\right] \geq \delta(\lambda)
\]
for some non-negligible function $\delta$. Then, on oracle access to $P^*$, the extractor $\mathcal{E}$ outputs $x' \in \mathbb{Z}_p$ such that $g^{x'} = w$ except with negligible probability:
\[
\Pr\left[
x' \leftarrow \mathcal{E}^{P^*}(g,w) \,\wedge\, g^{x'} = w
\right] \geq \delta(\lambda) - \mathsf{negl}(\lambda).
\]
Intuitively, any prover that makes $\mathsf{SPoSS.Verify}$ accept with non-negligible probability must ``know'' the discrete logarithm of $w$.

\paragraph*{Zero-knowledge.}
For any adversary that corrupts at most one of the two verifiers, there exists a probabilistic polynomial-time simulator $\mathcal{S}$ such that the adversary's view in a real SPoSS execution is computationally indistinguishable from its view in an ideal execution generated by $\mathcal{S}$, given only $w = g^x$ and oracle access to $H$. Formally, for each $I \subseteq \{1,2\}$ with $|I| = 1$, let
\[
\mathsf{Real}_I(x) := \bigl\{ (\pi_i,\tau_i) \mid i \in I \bigr\}
\]
be the joint distribution of proof and audit tokens observed by verifier(s) in $I$ in an honest execution on input $x$, and let
\[
\mathsf{Sim}_I(w) := \mathcal{S}(g,w,I)
\]
be the simulator's output distribution on input $w = g^x$. Then for all PPT distinguishers $\mathcal{D}$ we have
\[
\left|
\Pr\bigl[ \mathcal{D}(\mathsf{Real}_I(x)) = 1 \bigr]
- 
\Pr\bigl[ \mathcal{D}(\mathsf{Sim}_I(w)) = 1 \bigr]
\right|
\leq \mathsf{negl}(\lambda).
\]
In words, a coalition of fewer than two verifiers learns nothing about $x$ beyond the group element $w = g^x$ itself.

\end{document}